\documentclass[12pt]{article}

\usepackage{graphicx,psfrag,mathrsfs,bbm,amssymb,here,amsmath,cite}

\newcommand{\be}{\begin{equation}}
\newcommand{\ee}{\end{equation}}

\newcommand{\msp}{\;\;\;\;\;}
\newcommand{\boldm}[1]{\mbox{\boldmath ${#1}$}}

\newcommand{\lf}{\left}
\newcommand{\rg}{\right}

\newcommand{\ttilde}{\widetilde}
\newcommand{\hhat}{\widehat}
\newcommand{\bbar}{\overline}

\graphicspath{{figures/}}
\DeclareGraphicsExtensions{.eps.gz,.eps,.ps,.ps.gz}

\begin{document}                            

%%%%%%%%%%%%%%%%%%%%%%%%%%%%%%%%%%%%%%%%%%%%%%%%%%%%%%%%%%%%%%%%%
%%%%%%%%%%%%%%%%%%%%%%%%%%%%%%%%%%%%%%%%%%%%%%%%%%%%%%%%%%%%%%%%%
\pagestyle{empty}
\begin{flushright}
  LC-TH-2005-008 \\
  HD-THEP-05-19 \\
  hep-ph/0508133 
\end{flushright}

\vspace{\baselineskip}
 
\begin{center}
\textbf{\LARGE Effective-Lagrangian approach \\[0.3em]
        to \boldm{\gamma \gamma \rightarrow WW};\\[0.3em]
II: Results and comparison with \boldm{e^+e^- \rightarrow WW} \\}
\vspace{4\baselineskip}
{\sc O.~Nachtmann
\footnote{email:
O.Nachtmann@thphys.uni-heidelberg.de}, 
F.~Nagel
\footnote{email:
F.Nagel@thphys.uni-heidelberg.de}, 
M.~Pospischil
\footnote{Now at
CNRS UPR 2191, 1 Avenue de la Terrasse, F-91198 Gif-sur-Yvette,
France,\\\hspace*{.64cm}email: Martin.Pospischil@iaf.cnrs-gif.fr} 
and
A.~Utermann
\footnote{email:
A.Utermann@thphys.uni-heidelberg.de}
}\\
\vspace{1\baselineskip}
\textit{Institut f\"ur Theoretische Physik, Philosophenweg 16, D-69120
Heidelberg, Germany}\\
\vspace{2\baselineskip}
\textbf{Abstract}\\
\vspace{1\baselineskip}
\parbox{0.9\textwidth}{
We present a study of anomalous electroweak gauge-boson couplings
which can be measured in $e^+e^-$ and $\gamma\gamma$ collisions at
a future linear collider like ILC. We consider the gauge-boson
sector of a locally $SU(2)\times U(1)$ invariant effective
Lagrangian with ten dimension-six operators added to the
Lagrangian of the Standard Model. These operators induce anomalous
three- and four-gauge-boson couplings and an anomalous
\mbox{$\gamma \gamma H$}~coupling. We calculate the reachable
sensitivity for the measurement of the anomalous couplings in
$\gamma\gamma\to W W$. We compare these results with the
reachable precision in the reaction $e^+e^-\to WW$ on the
one hand and with the bounds that one can get from high-precision
observables in $Z$ decays on the other hand. We show that one
needs both the $e^+e^-$ and the $\gamma\gamma$ modes at an ILC to
constrain the largest possible number of anomalous couplings and
that the Giga-$Z$ mode offers the best sensitivity for certain
anomalous couplings.}
\end{center}
\vspace{\baselineskip}

\pagebreak

\pagestyle{plain}

\tableofcontents

\pagebreak

\section{Introduction}
\label{sec-intro} \setcounter{equation}{0}

A future linear electron-positron collider ILC with c.m. energies
of 500 GeV and more offers excellent possibilities for high
precision studies of the Standard Model (SM) of particle physics
\cite{Richard:2001qm,Ellis:1998wx,Badelek:2001xb,Burkhardt:2002vh}.
Particularly interesting is the electroweak gauge-boson sector
where the SM couplings are fixed by the requirement of
renormalisability. Deviations from the SM values for these
gauge-boson couplings would be a clear sign of new physics. A
comprehensive study of gauge-boson couplings is, therefore, highly
desirable. To this end all options for the ILC, in particular
$e^+e^-$ and $\gamma\gamma$ collisions, have to be considered.

In this paper we continue to explore and summarise the potential
of a $\gamma\gamma$ collider---especially in comparison to the
$e^+e^-$ mode---for constraining anomalous electroweak gauge-boson
couplings $h_i$ in an effective Lagrangian,
\begin{equation}
\label{eq-Leff}
\mathscr{L}_{\rm eff} = \mathscr{L}_{0} + \mathscr{L}_{2}\,,
\end{equation}
which respects the SM gauge symmetry group \mbox{$SU(3) \times
SU(2) \times U(1)$} and contains only the SM fields. Here
$\mathscr{L}_{0}$ is the Lagrangian of the SM and
\begin{equation}
\begin{split}
\mathscr{L}_{2} & = 
\Big(h_W O_W +
h_{\tilde{W}} O_{\tilde{W}} + h_{\varphi W} O_{\varphi W} + h_{\varphi
\tilde{W}} O_{\varphi \tilde{W}} + h_{\varphi B} O_{\varphi B} +
h_{\varphi \tilde{B}} O_{\varphi \tilde{B}} \\
\label{eq-Leff2}
 & \qquad +
h_{W\! B} O_{W\! B} +
h_{\tilde{W}\! B} O_{\tilde{W}\! B} +
h_{\varphi}^{(1)} O_{\varphi}^{(1)} +
h_{\varphi}^{(3)} O_{\varphi}^{(3)}\Big)/v^2\,,
\end{split}
\end{equation}
contains all dimension-six operators $O_i$ that either consist
only of electroweak gauge-boson fields ($W,B$ or
$\tilde{W},\tilde{B}$ respectively) or that contain both
gauge-boson fields and the SM-Higgs field ($\varphi$). For the
detailed definition of the operators in $\mathscr{L}_0$ and
$\mathscr{L}_2$ see App.~A in \cite{part1} and for a general
discussion of operators with dimension higher than four see
\cite{Buchmuller:1985jz,Leung:1984ni} and references therein.

In (\ref{eq-Leff2}) \mbox{$v \approx 246$}~GeV denotes the vacuum
expectation value of the SM-Higgs-boson field. The $h_i$ are the
anomalous couplings which are to be measured at the ILC. Since we
introduce the factor $1/v^2$ in $\mathscr{L}_{2}$ the couplings
$h_i$ are dimensionless. From a measurement or from bounds of a
$h_i$ one can estimate a lower bound on the scale of new physics as
\begin{equation}
\Lambda =\frac{v}{\sqrt{|h_i|+\delta h_i}}\,.
\end{equation}
Obviously, this approach is well suited to study effects of physics
beyond the Standard Model~(SM) at a future \mbox{$e^+e^-$} linear
collider~(LC) with a design like TESLA~\cite{Richard:2001qm} or
CLIC~\cite{Ellis:1998wx} in a model-independent way. A $\gamma\gamma$
collider---where two high-energy photons are obtained through Compton
backscattering of laser photons off high-energy electrons--- extends
the physics potential of a future LC substantially. Such a
photon-collider option has been discussed for example for
\mbox{$e^+e^-$}~machines like TESLA~\cite{Badelek:2001xb} or
CLIC~\cite{Burkhardt:2002vh}.

It is particularly interesting to study the rich phenomenology induced
by the Lagrangian~(\ref{eq-Leff}) at a future~LC both in the
high-energy \mbox{$e^+ e^-$} and~\mbox{$\gamma \gamma$} modes, and in
the Giga-$Z$ mode. In a preceding work~\cite{Nachtmann:2004ug} the
gauge-boson sector of such a Lagrangian and its implications for
$e^+e^-\to WW$ and for precision observables measured at the $Z$ pole
were studied. See also \cite{Hagiwara:1992eh,Monig:2003cm} and
references therein. In \cite{part1} we calculated the amplitudes for
the process $\gamma\gamma\to WW$ induced by the anomalous terms in the
Lagrangian. Also the relation to other approaches
\cite{Tupper:1980bw,Belanger:1992qi,Banin:1998ap,Marfin:2003jg,Bozovic-Jelisavcic:2002ta} 
to anomalous couplings, like the use of form factors, was discussed,
see \cite{part1}.

In this work we study, in the framework of the
Lagrangian~(\ref{eq-Leff}), the process \mbox{$\gamma \gamma
\rightarrow WW \rightarrow 4$}~fermions. In the reaction
\mbox{$\gamma \gamma \rightarrow WW$} anomalous contributions to
the \mbox{$\gamma WW$}, \mbox{$\gamma \gamma WW$} and
\mbox{$\gamma \gamma H$} vertices can be studied. In particular we
compare the results for the photon collider with those obtained
in~\cite{Nachtmann:2004ug} for the reaction
\mbox{$e^+e^-\rightarrow WW$} in order to see which anomalous
couplings can be measured best in which collider mode.

Our paper is organised as follows: In Sect.~\ref{sec-lep} we
summarise the results from~\cite{Nachtmann:2004ug} required here
and give the bounds on the anomalous couplings that are calculated
from present data. In Sect.~\ref{sec-cross} we review the
differential cross section for the process \mbox{$\gamma \gamma
\rightarrow WW$} with fixed c.m.~energy of the two-photon system
which was derived in the companion paper \cite{part1}.  In
Sect.~\ref{sec-opti} we review briefly the concept of the optimal
observables. In Sect.~\ref{secnum} we use the optimal observables to
calculate the reachable sensitivity to the anomalous couplings in
$\gamma\gamma\to WW$ at a future ILC. We compare these results with
the sensitivity reachable in the $e^+e^-$ mode both from $e^+e^-\to
WW$ and from $Z$ production. Our conclusions are presented in
Sect.~\ref{sec-conc}.
%%%%%%%%%%%%%%%%%%%%%%%%%%%%%%%%%%%%%%%%%%%%%%%%%%%%%%%%%%%%%%%%%
%%%%%%%%%%%%%%%%%%%%%%%%%%%%%%%%%%%%%%%%%%%%%%%%%%%%%%%%%%%%%%%%%

\section{Preliminaries and present constraints}
\label{sec-lep}
\setcounter{equation}{0}

In this section we summarise the results from~\cite{Nachtmann:2004ug}
that are required in this paper. We also give the present bounds on
the~$h_i$ as calculated in~\cite{Nachtmann:2004ug} from LEP and
SLC~results. The relation to other work on anomalous electroweak
gauge-boson couplings
\cite{Bozovic-Jelisavcic:2002ta,Gaemers:1978hg,Kuss:1997mf,
Diehl:2002nj,Diehl:1993br,Diehl:1997ft,Berends:1997av} is discussed in
\cite{Nachtmann:2004ug}.

After electroweak symmetry breaking some operators
of~$\mathscr{L}_{2}$ lead to new three- and four-gauge-boson
interactions, to \mbox{$\gamma \gamma H$}~interactions, and some
contribute to the diagonal and off-diagonal kinetic terms of the
gauge bosons and of the Higgs boson as well as to the mass terms
of the $W$ and $Z$~bosons. Thus one first has to identify the
physical fields~$A$, $Z$, $W^{\pm}$ and~$H$.  As explained
in~\cite{Nachtmann:2004ug} this requires a renormalisation of the
$W$-boson and of the Higgs-boson fields. Furthermore, the kinetic
matrix of the neutral gauge bosons has to be transformed into the
unit matrix while keeping their mass matrix diagonal in order to
obtain propagators of the standard form.  The full
Lagrangian~(\ref{eq-Leff}) is then expressed in terms of the
physical fields~$A$, $Z$, $W^{\pm}$ and~$H$.  In this procedure
the neutral- and charged-current interactions are modified
although no anomalous fermion-gauge-boson-interaction term is
added explicitly.  Therefore the Lagrangian~(\ref{eq-Leff}) leads
to a rich phenomenology to be probed at a future~LC both in the
high-energy \mbox{$e^+ e^-$} and~\mbox{$\gamma \gamma$} modes, and
in the Giga-$Z$ mode.

We now recall the input parameter schemes introduced
in~\cite{Nachtmann:2004ug}.  In the Lagrangian (\ref{eq-Leff}) the
Higgs potential is that of the~SM, cf.\ App.~A of \cite{part1}. It
contains the two parameters $\mu^2$ and~$\lambda$.  They can be
expressed in terms of the vacuum expectation value~$v$ and the
Higgs-boson mass~$m_H$.  Then the effective Lagrangian contains
the three electroweak parameters $g$, $g^\prime$ and~$v$ where $g$
and $g^\prime$ are the $SU(2)$ and $U(1)$ coupling constants.
Apart from that it contains the mass~$m_H$ of the Higgs boson,
nine fermion masses (neglecting neutrino masses), four parameters
of the CKM~matrix, and ten anomalous couplings~$h_i$.
In~\cite{Nachtmann:2004ug} two schemes, $P_Z$ and~$P_W$, were
introduced that include, instead of $g$, $g^\prime$ and~$v$, the
following electroweak parameters:
\begin{equation}
\label{eq-schemes}
P_Z:\;\; \alpha(m_Z),\, G_{\rm F},\, m_Z;\msp \msp \msp P_W:\;\; \alpha(m_Z),\,
G_{\rm F},\, m_W\,.
\end{equation}
For the list of the other parameters, which are identical in both
schemes, see Table 3 of \cite{Nachtmann:2004ug}. Here \mbox{$\alpha
(m_Z)$} is the fine-structure constant at the $Z$~scale, $G_{\rm F}$
is the Fermi constant, and $m_Z$ and $m_W$ are the masses of the $Z$
and the $W$~bosons, respectively.  All constants of the
Lagrangian~(\ref{eq-Leff}) can then be expressed in terms of one of
the two parameter sets~(\ref{eq-schemes}), the other SM parameters and
the anomalous couplings~$h_i$.  Expressing the
Lagrangian~(\ref{eq-Leff}) in terms of the physical gauge-boson
fields, the physical Higgs-boson field~$H$, and the parameters of
either~$P_Z$ or~$P_W$, $\mathscr{L}_{\rm eff}$ shows a non-linear
dependence on the anomalous couplings~$h_i$, although the original
expression~(\ref{eq-Leff}) is linear in the~$h_i$.  This
non-linearity stems from the renormalisation of the $W$ and the
Higgs~fields, and from the simultaneous diagonalisation in the
neutral-gauge-boson sector as well as from expressing all constants in
terms of the above input-parameter sets. We then expand the full
Lagrangian~(\ref{eq-Leff}) in the~$h_i$ and drop all terms of second
or higher order in the~$h_i$. That is, we keep only the leading order
terms in $(v/\Lambda)^2$. Throughout this paper we
consider the Lagrangian~(\ref{eq-Leff}) after this reparametrisation
and linearisation unless otherwise stated.  Of course, the resulting
expressions depend on whether we choose $P_W$ or~$P_Z$ as parameter
scheme.

As shown in~\cite{Nachtmann:2004ug} in linear order in the $h_i$
the neutral-current boson-fermion interactions are modified by
anomalous couplings in both schemes, the charged-current
interactions are changed only in the scheme~$P_Z$ but in a
universal way for all fermions.  Furthermore, the $W$- ($Z$-)boson
mass changes in the scheme~$P_Z$~($P_W$) in order~$h_i$.  Due to
these modifications of the neutral- and charged-current
interactions and of the boson-masses, bounds on the~$h_i$ from
electroweak precision measurements at~LEP and~SLC can be derived
as done in this framework in~\cite{Nachtmann:2004ug}.  There the
scheme~$P_Z$ is used for this purpose since $m_Z$ is known more
precisely than~$m_W$.  Without data from direct measurements of
triple-gauge-boson couplings~(TGCs) these bounds are of
order~10$^{-3}$ for $h_{W\! B}$ and~$h_{\varphi}^{(3)}$. Moreover,
a number of gauge-boson self-interactions and
gauge-boson-Higgs-interactions are modified in order~$h_i$, see
Tab.~\ref{tab:vertices}.  Note that we only list those vertices
there that are relevant either in this paper or for the
observables considered in~\cite{Nachtmann:2004ug}.
\begin{table}
\centering
\begin{tabular}{|r|ccccccccccc|}
\hline
 &&&&&&&&&&&\\[-.45cm]
 & SM & $h_W$ & $h_{\tilde{W}}$ & $h_{\varphi W}$ & $h_{\varphi \tilde{W}}$
 & $h_{\varphi B}$ & $h_{\varphi \tilde{B}}$ & $h_{W\! B}$ &
 $h_{\tilde{W}\! B}$ & $h_{\varphi}^{(1)}$ &
 $h_{\varphi}^{(3)}$\\[.1cm]
\hline
 &&&&&&&&&&&\\[-.43cm]
$\gamma WW$ & $\surd$ & $\surd$ & $\surd$ & & & & &
 $\surd$ & $\surd$ & &\\
$ZWW$ & $\surd$ & $\surd$ & $\surd$ & & & & & $\surd$ & $\surd$ &  & $P_Z$\\
$\gamma \gamma WW$ & $\surd$ & $\surd$ & $\surd$ & & & & & & & &\\
$\gamma \gamma H$ & & & & $\surd$ & $\surd$ & $\surd$ & $\surd$ & $\surd$ &
 $\surd$ & &\\[.05cm]
\hline
\end{tabular}
\caption{\label{tab:vertices}Contributions of the SM Lagrangian and of
the anomalous operators~(\ref{eq-Leff}) to different vertices in
order~\mbox{$O(h)$}.  The coupling~$h_{\varphi}^{(3)}$ contributes to the
$ZWW$~vertex in the scheme~$P_Z$ but not in~$P_W$.}
\end{table}
Including direct measurements of~TGCs at LEP2, one further
$CP$~conserving coupling~($h_W$) and two $CP$~violating couplings
($h_{\tilde{W}}$, $h_{\tilde{W}\! B}$) are constrained in
\cite{Nachtmann:2004ug}.

We now summarise the numerical results for these anomalous couplings calculated
in~\cite{Nachtmann:2004ug} from data obtained at LEP1, SLC, LEP2 and
from further $W$-boson data.  These bounds were derived in the scheme
$P_Z$, see~(\ref{eq-schemes}).  The following numerical values for the input
parameters were used~\cite{Hagiwara:pw,:2002mc}:
\begin{align}
\label{eq-inputalpha}
1 / \alpha (m_Z) & =  128.95(5)\,, 
\\  \label{eq-inputgfermi}
G_{\rm F} & =  1.16639(1) \times 10^{-5}\;{\rm GeV}^{-2}\,,
\\  \label{eq-inputmz}
m_Z & =  91.1876(21)\;{\rm GeV}\,,
\end{align}
and in the $P_W$ scheme \cite{Hagiwara:pw},
\begin{equation}
    \label{eq-inputmw}
 m_W=80.423(39)\;\text{GeV}\,.
\end{equation}
Here as in \cite{Nachtmann:2004ug}, we use the following definition
of the effective leptonic weak mixing angle:
\begin{equation}
s_{\rm eff}^2 \equiv \sin^2 \theta^{\rm lept}_{\rm eff} = \frac{1}{4} \bigg(
1 - \frac{g_{\rm V}^{\ell}} {g_{\rm A}^{\ell}}\bigg)\,,
\end{equation}
where $g_{\rm V}^{\ell}$ and $g_{\rm A}^{\ell}$ are the vector and
axial-vector neutral-current couplings of the leptons, \mbox{$\ell
= e, \mu,\tau$}.  In the $P_Z$~scheme this quantity contains a
particular linear combination of the couplings $h_{W\! B}$
and~$h_{\varphi}^{(3)}$, see~(5.4) in~\cite{Nachtmann:2004ug}:
\begin{equation}
\label{sseff}
s_{\rm eff}^2 = \left(s_{\rm eff}^{\rm SM}\right)^2 \left(1 + 3.39\,h_{W \mskip
    -3mu B} + 0.71 \,h_{\varphi}^{(3)}\right)\,.
\end{equation}
A large number of $Z$-pole observables depends on the anomalous
couplings only through $s_{\rm eff}^2$, see table 16.1 of
\cite{:2002mc} and Sect. 5 of \cite{Nachtmann:2004ug}. Thus the
measured value of $s^2_\text{eff}$ determines bounds on the linear
combination of $h_{W \mskip -3mu B}$ and $h_{\varphi}^{(3)}$
occurring in (\ref{sseff}). The total width of the $Z$~boson,
$\Gamma_Z$, the mass~$m_W$ and width~$\Gamma_W$ of the $W$~boson
depend on these two anomalous couplings in a different way.
Therefore from the measured values of~$s_{\rm eff}^2$, $\Gamma_Z$,
$\sigma_{\rm had}^0$, $R_{\ell}^0$, $m_W$ and~$\Gamma_W$ bounds on
these two couplings individually are obtained in
\cite{Nachtmann:2004ug}. One further $CP$~conserving coupling,
$h_W$, enters the three-gauge-boson vertices~\mbox{$\gamma WW$}
and~\mbox{$ZWW$} and therefore can be constrained when considering
direct measurements of the~TGCs.  Altogether, from present data
the three $CP$~conserving couplings~$h_{W\! B}$,
$h_{\varphi}^{(3)}$ and~$h_W$ can be determined.  We list the
corresponding results from~\cite{Nachtmann:2004ug} in
Tab.~\ref{tab:couplep}.  Since the SM~predictions of these
observables depend on the mass~$m_H$ of the Higgs boson, the
bounds for the~$h_i$ are functions of~$m_H$, too.
\begin{table}
\centering
\begin{tabular}{|lr|ccc|c|rrr|}
\hline 
 \multicolumn{9}{|c|}{\rule[-2.5mm]{0mm}{7.5mm}$s_{\rm eff}^2$, $\Gamma_Z$, $\sigma^0_{\rm had}$,
 $R^0_{\ell}$, $m_W$, $\Gamma_W$, TGCs} 
\\ \hline
$m_H$& \hspace{-.5cm} & 120 GeV & 200 GeV & 500 GeV & $\delta h \times 10^3$
&&W($h$)&\\
\hline
 &&&&&&&&\\[-.45cm]
$h_W$ \hspace{-.5cm} & $\times 10^3$ & $-62.4$ & $-62.5$ & $-62.8$ & 36.3 &
\msp 1 & $-0.007$ & $0.008$\\
$h_{W\! B}$ \hspace{-.5cm} & $\times 10^3$ & $-0.06$ & $-0.22$ &
$-0.45$ & 0.79 & & 1 & $-0.88$\\
$h_{\varphi}^{(3)}$ \hspace{-.5cm} & $\times 10^3$ & $-1.15$ & $-1.86$ &
$-3.79$ & 2.39 & & & 1\\
\hline
\end{tabular}
\caption{\label{tab:couplep}Final results from \cite{Nachtmann:2004ug}
 for $CP$~conserving couplings in units of $10^{-3}$ for a Higgs mass
 of 120~GeV, 200~GeV and 500~GeV respectively. The anomalous couplings
 are extracted from existing electroweak precision data for the
 observables listed in the first row. The errors~\mbox{$\delta h$} and
 the correlations of the errors are independent of the Higgs mass
 within the given accuracy. The correlation matrix is given in the
 three right most columns.}
\end{table}

In addition, from the measurement of TGCs at LEP2 the following
values for two $CP$~violating couplings are obtained
in~\cite{Nachtmann:2004ug},
\begin{alignat}{2}
\label{eq-hwtilde}
h_{\tilde{W}} & = &\; 0.068 &\pm 0.081\,,\\
\label{eq-hwtildeb}
h_{\tilde{W}\! B} & = &\; 0.033 &\pm 0.084\,.
\end{alignat}

%%%%%%%%%%%%%%%%%%%%%%%%%%%%%%%%%%%%%%%%%%%%%%%%%%%%%%%%%%%%%%%%%
%%%%%%%%%%%%%%%%%%%%%%%%%%%%%%%%%%%%%%%%%%%%%%%%%%%%%%%%%%%%%%%%%

\section{The process \boldm{\gamma \gamma \rightarrow WW
\rightarrow} 4 fermions}
\label{sec-cross}
\setcounter{equation}{0}
\subsection{Fixed photon energies}
\label{secfixed}

\begin{figure}[h]
\centering
\includegraphics[totalheight=5cm]{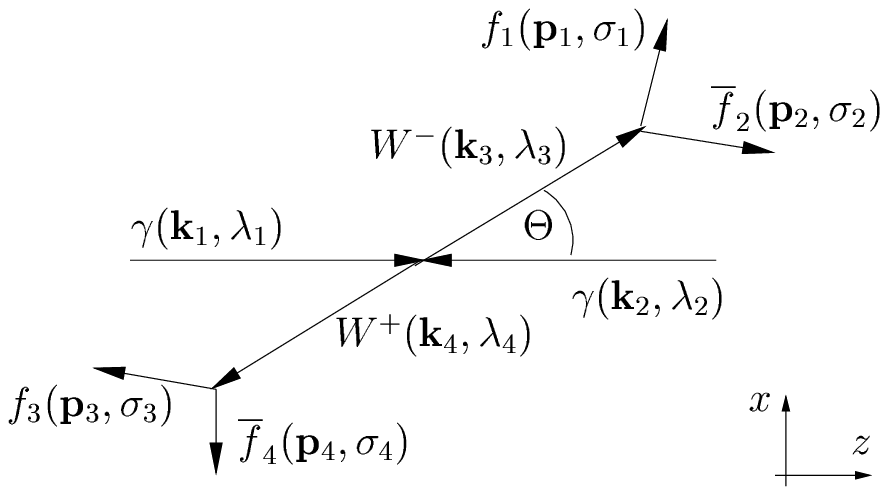}
\caption{\label{fig:conv}Conventions for particle momenta and helicities.}
\end{figure}

In this section we review briefly the differential cross section
for the process $\gamma\gamma\to WW\to 4\text{f}$ in the presence
of anomalous couplings. For more details see \cite{part1}. It is
essential here to use the $P_W$ scheme (\ref{eq-schemes}) since in
the $P_Z$ scheme the $h_i$ would modify the $W$~mass and therefore
the kinematics of the reaction, which is highly inconvenient. The
final-state fermions are leptons or quarks and we start with fixed
photon energies. The case where the initial photons are not
monochromatic but have Compton-energy spectra will be considered
in the following section. Our notation for particle momenta and
helicities is shown in Fig.~\ref{fig:conv}.
The production of the $W$~bosons is described in the \mbox{$\gamma
\gamma$ c.m.\ frame}.  Our coordinate axes are chosen such that
the \mbox{$WW$}-boson production takes place in the $x$-$z$ plane,
the photon momentum ${\bf k}_1$ points in the positive
$z$-direction and the unit vector in $y$-direction is given by
\mbox{$\boldm{\hat{e}}_y = ({\bf k}_1 \times {\bf k}_3)/|{\bf k}_1
\times {\bf k}_3|$}.

For unpolarised photons we obtain in the narrow-width approximation
for the $W$~bosons and considering all final-state fermions to be
massless
\begin{equation}
\label{eq-diffs}
\frac{{\rm d}\sigma}{{\rm d}\!\cos \Theta \; {\rm d}\!\cos \vartheta \;  {\rm
d}\varphi \; {\rm d}\!\cos \overline{\vartheta} \;  {\rm d}\overline{\varphi}}
= \frac{3\beta}{2^{13} \pi^3 s} \, B_{12} \, B_{34} \, \mathcal{P}^{\lambda_3
\lambda_4}_{\lambda^\prime_3 \lambda^\prime_4} \, \mathcal{D}^{\lambda_3}_{\lambda^\prime_3} \,
\overline{\mathcal{D}}^{\lambda_4}_{\lambda^\prime_4}\;.
\end{equation}
Here summation over repeated indices is implied and \mbox{$\beta
= (1-4 m_W^2 / s)^{1/2}$} is the velocity of each $W$~boson in the
\mbox{$\gamma \gamma$}~c.m.~frame.  The branching ratio for the
decay \mbox{$W \rightarrow f_i \overline{f}_j$} is denoted by
$B_{ij}$. The $W$~helicity states are defined in the coordinate
system shown in Fig.~\ref{fig:conv}. For the definition of the
polarisation vectors see App.~C in \cite{part1}. The polar angle
between the positive $z$-axis and the $W^-\!$~momentum is denoted
by~$\Theta$.  The cross section does not depend on the azimuthal
angle of the $W^-\!$~momentum due to rotational invariance. The
respective frames for the decay tensors are defined by a rotation
by $\Theta$ about the $y$-axis of the frame in Fig.~\ref{fig:conv}
such that the $W^-\!$ ($W^+$) momentum points in the new positive
(negative) $z$-direction and a subsequent rotation-free boost into
the c.m.~system of the corresponding $W$~boson.  The spherical
coordinates $\vartheta, \varphi$ and
$\overline{\vartheta},\overline{\varphi}$ are those of the $f_1$-
and $\overline{f}_4$-momentum directions, respectively.  The
$WW$-production and $W$-decay tensors in (\ref{eq-diffs}) are
given by
\begin{align}
\label{eq-prodpro} \mathcal{P}^{\lambda_3
\lambda_4}_{\lambda^\prime_3 \lambda^\prime_4} (\Theta) & =
\sum_{\lambda_1, \lambda_2} \mathcal{M}(\lambda_1, \lambda_2;
\lambda_3,
\lambda_4) \mathcal{M}^*(\lambda_1, \lambda_2; \lambda^\prime_3,\lambda^\prime_4)\,, \\
\mathcal{M}(\lambda_1,\lambda_2;\lambda_3,\lambda_4)&\equiv
\langle W^-({\bf k}_3, \lambda_3)\,W^+({\bf k}_4,\lambda_4)|\mathcal{T}|\gamma({\bf k_1},\lambda_1)\,\gamma({\bf k_2},\lambda_2)\rangle\,,\\[1ex]
\label{eq-decay12}
\mathcal{D}^{\lambda_3}_{\lambda^\prime_3}(\vartheta, \varphi) & =
l^{\phantom{*}}_{\lambda_3}
l^*_{\lambda^\prime_3}\;,\\	
\label{eq-decay34}
\overline{\mathcal{D}}^{\lambda_4}_{\lambda^\prime_4}(
\overline{\vartheta},\overline{\varphi}) & =
\overline{l}^{\phantom{*}}_{\lambda_4}
\overline{l}^{\,*}_{\lambda^\prime_4}\;,
\end{align}
where we have suppressed the phase-space variables on the right
hand side. The production amplitudes $\mathcal{M}$ and the
functions occurring in the decay tensors are listed in App.~D in
\cite{part1}.

To first order in the anomalous couplings the amplitudes
$\mathcal{M}$ are obtained from the SM diagrams, diagrams
containing one anomalous triple- or quartic-gauge-boson vertex and
the $s$-channel Higgs-boson exchange. The Feynman rules that are
necessary to compute these diagrams are listed in App.~B in
\cite{part1}. This gives
 \begin{equation}
 \label{eq-Mdecomp}
\mathcal{M} = \mathcal{M}_{\rm SM} + \sum_i h_i \mathcal{M}_i \; +
\; O(h^2)\,,
\end{equation}
where all particle momenta and helicities are suppressed.
$\mathcal{M}_{\rm SM}$ is the SM tree-level amplitude and \mbox{$i
= W, \tilde{W}, \varphi W, \varphi \tilde{W}, \varphi B, \varphi
\tilde{B}, W\! B, \tilde{W} \mskip -3mu B$}.  The
couplings~$h_{\varphi}^{(1)}$ and~$h_{\varphi}^{(3)}$ do not enter
the amplitudes~(\ref{eq-Mdecomp}) to first order, since the
related operators do not contribute to any anomalous gauge-boson
vertex (see Tab.~\ref{tab:vertices}).

Since the couplings \mbox{$h_i, i = W, \tilde{W}, \varphi W,
\varphi \tilde{W}, \varphi B, \varphi \tilde{B}, W\! B, \tilde{W}
\mskip -3mu B$} contribute to the differential cross section, one
could expect that these 8 couplings are measurable at a photon
collider. But this is not the case since some amplitudes
$\mathcal{M}_i$ are related in a trivial way. We find the following
relations between amplitudes, independently of helicities and momenta,
\begin{align}
 s_1^2\,\mathcal{M}_{\varphi B} &
=c_1^2\,\mathcal{M}_{\varphi W}\,,
\label{Mphirel}
\\
\label{Mphitilderel}
 s_1^2\,\mathcal{M}_{\varphi \tilde{B}} &
=c_1^2\,\mathcal{M}_{\varphi \tilde{W}}\,,
\end{align}
where
\begin{equation}
 s_1^2\equiv\frac{e^2}{4\,\sqrt{2}\,G_F\,m_W^2}\,,\qquad
c_1^2\equiv 1-s_1^2\,, \label{s1c1}
\end{equation}
are combinations of input parameters in the $P_W$ scheme. Hence the
corresponding four anomalous couplings do not appear in the amplitudes
in an independent way but only as linear combinations
\begin{align}
 h_{\varphi W\! B}
&\equiv s_1^2\,h_{\varphi W}
      + c_1^2\,h_{\varphi B}\,,
\label{eq-hphiWB}
\\
\label{eq-hphiWtildeB}
h_{\varphi \tilde{W}\!\tilde{B}}
&\equiv s_1^2\,h_{\varphi \tilde{W}}
       +c_1^2\,h_{\varphi \tilde{B}}\,.
\end{align}
 We conclude that the process \mbox{$\gamma
\gamma \rightarrow WW$} is sensitive to the anomalous couplings $h_W$,
$h_{\tilde{W}}$, $h_{\varphi W\! B}$, $h_{\varphi \tilde{W}\!\tilde{B}}$, $h_{W
\mskip -3mu B}$ and~$h_{\tilde{W}\! B}$. The
couplings $h_\varphi^{(1)}, h_\varphi^{(3)}$ and the orthogonal
combinations to (\ref{eq-hphiWB}) and (\ref{eq-hphiWtildeB}), that is
\begin{align}
  h^\prime_{\varphi W\! B} &
=c_1^2 h_{\varphi W} -s_1^2 h_{\varphi B} \,,
\label{eq-hphiWBprime}
\\
\label{eq-hphiWtildeBprime} h^\prime_{\varphi
\tilde{W}\!\tilde{B}} & =  c_1^2 h_{\varphi \tilde{W}} -s_1^2
h_{\varphi\tilde{B}}\,,
\end{align}
do not enter in the expressions for the amplitudes of
$\gamma\gamma\to WW$  due to (\ref{Mphirel}) and (\ref{Mphitilderel}).

%%%%%%%%%%%%%%%%%%%%%%%%%%%%%%%%%%%%%%%%%%%%%%%%%%%%%%%%%%%%%%%%%
%%%%%%%%%%%%%%%%%%%%%%%%%%%%%%%%%%%%%%%%%%%%%%%%%%%%%%%%%%%%%%%%%

\subsection{Photons at a $\gamma\gamma$ collider}
\label{ssecxs2}

In the last section we discussed the differential cross section of
the process $\gamma\gamma\to WW\to 4\text{f}$ for fixed
$\gamma\gamma$ c.m. energy $\sqrt{s}$. However, at a real
$\gamma\gamma$ collider the photons do not have fixed energies in
the laboratory system (LS) but they have a rather wide energy
distribution. We now consider unpolarised photons whose energy is distributed
according to a Compton spectrum \cite{Ginzburg:1981vm}. This is
still not completely realistic but good enough for our purposes.

We consider two beams of a collider where electrons of energy $E_e$
(in the LS) scatter on laser photons of energy $\omega$ to give
high-energy photons by Compton scattering. According to (30a) in
\cite{Ginzburg:1981vm} the $\gamma\gamma$ luminosity spectrum is given
by
\begin{equation}
\text{d}L_{\gamma\gamma}=k^2\,L_{ee}\,
f\lf(x,\frac{E_1}{E_e}\rg)\,f\lf(x,\frac{E_2}{E_e}\rg) \;\frac{\text{d}E_1}{E_e}\,\frac{\text{d}E_2}{E_e}\,,
\label{compton}
\end{equation}
where $E_1$ and $E_2$ are the energies of the two photons in the
LS, $L_{ee}$ is the luminosity of the $e^+e^-$ collider and
$k$ is the conversion factor for the $\gamma$ production, see
\cite{Ginzburg:1981vm} for further details. The parameter $x$ is
given by
\begin{equation}
x = \frac{4\,E_e\,\omega}{m_e^2}\,\cos^2\frac{\alpha}{2}\,,
\label{defx}
\end{equation}
where $m_e$ is the electron mass, $\omega$ is the energy of the
laser photons and $\alpha$ is the angle between the incoming
electron and the laser photon. Throughout this paper, we use $x=4.6$.  The
energy spectrum of the scattered photons is,
cf. (6a) in
\cite{Ginzburg:1981vm}:
\begin{equation}
f(x,y) = \lf(\ln x + \frac{1}{2}\rg)^{-1}
\lf[1-y+\frac{1}{1-y}-\frac{4y}{x(1-y)}+
\frac{4y^2}{x^2(1-y)^2}\rg]\,. \label{fdef}
\end{equation}
Because we expressed the differential cross section in
Sect.~\ref{secfixed} as function of the squared $\gamma\gamma$
c.m. energy $s =4E_1E_2$, it is convenient to express the spectrum
(\ref{compton}) in terms of $s$ and $E_1$ instead of $E_1$ and
$E_2$. Using (\ref{compton}), one gets,
\begin{equation}
\text{d}L_{\gamma\gamma}= \frac{k^2\,L_{ee}}{2\,E_e^2}
 \frac{\sqrt{s}}{E_1}\,
f\lf(x,\frac{E_1}{E_e}\rg)\, f\lf(x,\frac{s}{4\,E_1\,E_e}\rg)
\;\text{d}E_1\,\text{d}\sqrt{s} \,. \label{compton2}
\end{equation}
For the unpolarised differential cross section of the process
$ee\to\gamma\gamma\to WW\to 4\:\text{f}$ induced by
Compton backscattering we hence obtain:
\begin{align}
\frac{\text{d}\sigma_{ee\to\gamma\gamma\to WW\to
4\:\text{f}}}{\text{d}\sqrt{s}\:\text{d}E_1\:\text{d}\phi}
&=\frac{1}{L_{ee}}\,
\frac{\text{d}L_{\gamma\gamma}}{\text{d}\sqrt{s}\:\text{d}E_1}\;
\frac{\text{d}\sigma\lf(\sqrt{s}\rg)}{\text{d}\phi} =
\frac{k^2}{2\,E_e^2}\:\ttilde{S}\,,
\label{cross-comp}
\\
\ttilde{S}&\equiv
 \frac{\sqrt{s}}{E_1}\,
f\lf(x,\frac{E_1}{E_e}\rg)\, f\lf(x,\frac{s}{4\,E_1\,E_e}\rg)
\;\frac{\text{d}\sigma\lf(\sqrt{s}\rg)}{\text{d}\phi}\,,
\label{stilde}
\end{align}
where
\begin{equation}
 \phi=(\Theta, \vartheta, \varphi, \bbar{\vartheta}, \bbar{\varphi})
 \label{phi}
\end{equation}
stands for the set of five phase space variables defined in Fig.~\ref{fig:conv}
and $\text{d}\sigma\lf(\sqrt{s}\rg)/\text{d}\phi$ is the differential
cross section (\ref{eq-diffs}) for fixed $\gamma\gamma$ c.m.
energy $\sqrt{s}$. 

\subsection{Kinematics}
\label{ssecxs3}
It is now easy to see that the final state in the reaction
$\gamma\gamma\to WW\to 4\,\text{fermions}$ at a photon collider is uniquely specified by
the 7 variables
\begin{equation}
\chi\equiv\lf(\sqrt{s},E_1,\phi\rg)\,, \label{chi}
\end{equation}
considering unpolarised photons and summing over the helicities of the
final fermions. We would like to determine to which extent the
variables $\chi$ can be reconstructed in an experiment if we consider
the case of one $W$ decaying leptonically and one to a quark-antiquark
pair, $q\bar{q}$, that is to two jets.  We suppose that the jets are
not tagged as $q$ or $\bar{q}$ jet and that, therefore, the
(anti)quark cannot be associated to one of the two jets. We assume that the
following variables can be measured:
\begin{equation}
 \xi\equiv\lf(k_{W,x},k_{W,z},\hhat{\bf k}_\text{jet},{\bf k}_\ell\rg)\,.
 \label{xi}
\end{equation}
Here $k_{W,x}$ and $k_{W,z}$ are the $x$- and $z$-components of the
momentum in the LS of the $W$ boson that decays into two
jets. Furthermore, ${\bf\hhat{k}_\text{jet}}$ is the momentum
direction of one of the jets in the rest frame of this $W$ and ${\bf
k}_\ell$ is the momentum of the charged lepton from the decay of the
other $W$ in the LS. In our coordinates we have $k_{W,y}=0$, see
Fig.~\ref{fig:conv}. Thus 7 quantities are measurable on which the
cross section depends in a non-trivial way. From counting of variables
we therefore conclude that the full set of variables $\chi$ may be
reconstructible. However, there can be ambiguities, that is for events
with certain measured kinematic variables  $\xi$ there may correspond two or more
final states $\chi_k$ with $k=1,2,\ldots$ that cannot be
distinguished. Two different ambiguities occur in our reaction. The
first one is due to the fact that the neutrino momentum cannot be
directly measured, the second one occurs because the jet charges are
not tagged. In App.~\ref{sec-kine} we take a closer look at these
ambiguities.

%%%%%%%%%%%%%%%%%%%%%%%%%%%%%%%%%%%%%%%%%%%%%%%%%%%%%%%%%%%%%%%

\section{Optimal observables}
\setcounter{equation}{0}
\label{sec-opti}

In Sect.~\ref{sec-lep} we saw that the operators $O_{W\! B}$ and
$O^{(3)}_\varphi$ have an impact on electroweak precision observables
measured at LEP and SLD, whereas the operators $O_{W}, O_{\tilde{W}},
O_{W\!B}$ and $O_{\tilde{W}\!B}$ affect the $W$-pair production in
$e^+e^-$ collisions. According to the results of Sect.~\ref{sec-cross}
we can constrain even more couplings in the process $\gamma\gamma\to
WW\to 4\;\text{fermions}$, that is the linear combinations $h_{\varphi
W\! B}$ (\ref{eq-hphiWB}) and $h_{\varphi\tilde{W}\!\tilde{B}}$
(\ref{eq-hphiWtildeB}) which enter in the now accessible anomalous
$\gamma\gamma H$ vertex, see Tab.~\ref{tab:vertices}. To compute the
maximum sensitivity of the normalised differential distribution to the
anomalous couplings we use optimal observables
\cite{Atwood:1991ka,Diehl:1993br,Diehl:1997ft}. This method has been applied to
analyses of triple gauge couplings in the reaction $e^+e^-\to WW$ in
\cite{Diehl:2002nj,Diehl:1993br,Diehl:1997ft}
which takes into account all statistical correlations of the errors
on the couplings. We summarise below the general properties of
optimal observables, see
\cite{Nachtmann:2004fy} for details. 

\noindent
 In an experiment one measures the differential cross section
\begin{equation}
 S(\xi) = \text{d}\sigma/\text{d}\xi\,,\label{sxi}
\end{equation}
where $\xi$ denotes the set of all measured phase space variables.
Expanding $S$ in the anomalous couplings one can write
\begin{equation}
S(\xi) = S_0(\xi) + \sum_i S_i(\xi)\,h_i + O(h^2)\,, \label{sevo}
\end{equation}
where $S_0(\xi)$ is the tree-level cross section in the SM. One way
to extract the anomalous couplings from the measured distribution
(\ref{sevo}) is to look for a suitable set of observables
$\mathcal{O}_i(\xi)$ whose expectation values
\begin{equation}
E[\mathcal{O}_i] = \frac{1}{\sigma} \int
\text{d}\xi\;S(\xi)\,\mathcal{O}_i(\xi) \label{eoi}
\end{equation}
 are sensitive to the dependence of $S$ on the couplings $h_i$.
Here, we will use the following observables,
\begin{equation}
\mathcal{O}_i(\xi) =\frac{S_i(\xi)}{S_0(\xi)}\,. \label{oi}
\end{equation}
These observables are optimal in the sense that for $h_i \to 0$ the
errors on the couplings extracted from them are as small as they can
be for a given probability distribution, see
\cite{Diehl:1993br,Diehl:1997ft,Nachtmann:2004fy}. Using this set of observables
the resulting covariance matrix is,
\begin{align}
\label{vh}
  V(h) &= \frac{1}{N} c^{-1}+O(h)\,,
\\[1ex]
\label{cij}
  c_{ij} &=
\frac{\int\text{d}\xi\:(S_i(\xi)\,S_j(\xi))/S_0(\xi)}
{\int\text{d}\xi\:S_0(\xi)}
-
\frac{\int\text{d}\xi\:S_i(\xi)\;\int\text{d}\xi\:S_j(\xi)}
{\lf(\int\text{d}\xi\:S_0(\xi)\rg)^2}
\,.
\end{align}
Apart from being useful for actual experimental analyses, the
observables (\ref{oi}) thus provide insight into the sensitivity that
is at best attainable by {\em any} method, given a certain
process---described by a differential cross section $S(\xi)$---and
specified experimental conditions. We further note that phase space
cuts, as well as detector efficiency and acceptance have no influence
on the observables being "optimal" in the above sense, since their
effects drop out in the ratio (\ref{oi}). This is not the case for
detector resolution effects, but the observables (\ref{oi}) are still
close to optimal if such effects do not significantly distort the
differential distributions $S_i$ and $S_0$ (or tend to cancel in their
ratio). To the extent that they are taken into account in the data
analysis, none of these experimental effects will bias the estimators.

In the present work we use the method of optimal observables in the
linear approximation valid for small anomalous couplings. But we
emphasise that in \cite{Diehl:1997ft} the method has been extended to
the fully non-linear case where one makes no a priori assumptions on
the size of anomalous couplings.

With the differential cross section (\ref{eq-diffs}) and the
covariance matrix (\ref{vh}) in leading order of the anomalous
couplings we have the basic ingredients at hand to calculate the
sensitivities $\delta h_i=\sqrt{V_{ii}}$. 

Now we construct the functions $S_0(\xi)$ and $S_i(\xi)$ needed for
the optimal observables (\ref{oi}). The reconstruction ambiguities
discussed in Sect.~\ref{ssecxs3} and App.~\ref{sec-kine} introduce
some slight complications. We start from a particular set of
phase-space variables~$\chi$ that specifies the final state uniquely. In
our case this set is given by (\ref{chi}). The differential cross
section in terms of these variables is
\begin{equation}
T(\chi) \equiv {\rm d}\sigma / {\rm d}\chi\,.
\end{equation}
The cross section for the measurable set of variables $\xi$
(\ref{xi}), is then 
\begin{equation}
\label{eq-tdef}
S(\xi) = \int {\rm d}\chi \; \delta (F(\chi) - \xi) \, T(\chi)\,.
\end{equation}
The function~$F$, expressing the relation of~$\xi$ to~$\chi$, may take the same
value for different values of~$\chi$, that is for a given~$\xi$ the equation
\begin{equation}
\label{eq-solf}
F(\chi) = \xi
\end{equation}
may have several solutions~\mbox{$\chi_k\equiv\chi_k(\xi)$} with
\mbox{$k = 1,2,\ldots$}.  In general, the number of solutions
to~(\ref{eq-solf}) may vary with~$\xi$.  If $\xi$ are the
coordinates that can be measured of an event~$\chi$, the set of final
states~$\chi_k$ consists of $\chi$ itself as well as all final states
that cannot be distinguished from~$\chi$ by a measurement of~$\xi$.
Coming back to~(\ref{eq-tdef}) we have
\begin{equation}
\label{eq-t2}
S (\xi) = \sum_k |J_k|^{-1} \; T(\chi_k(\xi))
\end{equation}
where
\begin{equation}
\label{eq-jacob}
J_k \equiv \det \frac{\partial F}{\partial \chi} (\chi_k(\xi))
\end{equation}
is the Jacobian determinant taken at point~$\chi_k$.

The cross section $S(\xi)$ in (\ref{eq-t2}) is to be used for the
construction of the optimal observables according to (\ref{sxi}) -
(\ref{cij}). The sums which are generally occurring in (\ref{eq-t2})
with the number of terms in the sum depending on $\xi$ must be
adequately treated in the integrations of (\ref{eoi}) and
(\ref{cij}). See \cite{Nachtmann:2004fy} for the details.

We now apply these considerations to the reaction $\gamma\gamma\to WW
\to 4\;\text{fermions}$. At a photon collider, the photons have a
nontrivial energy spectrum as described in Sect.~\ref{ssecxs2}.  The
final state is specified uniquely by the variables $\chi$, see
(\ref{chi}). If we assume that the variables $\xi$ in (\ref{xi}) are
measurable we obtain a two-fold ambiguity in the reconstruction of
$\xi$ for part of the phase space and a four-fold one for another
part, see Sect.~\ref{ssecxs3} and App.~\ref{sec-kine}. We must now
calculate $F(\chi)$ (\ref{eq-solf}) and the Jacobian
(\ref{eq-jacob}). 

Suppose first that the $W^+$ decays into leptons and the $W^-$
hadronically.  After performing the appropriate boosts and rotations
between the reference frames defined in Sect.~\ref{sec-cross} and the
LS we obtain
\begin{align}
k_{W,x} &= m_W\,\gamma\,\beta\,\sin\Theta\,, \label{kwx}
\\[0.6ex]
k_{W,z} &= m_W\,\gamma (b_- + b_+\,\beta\,\cos\Theta) \,,
\\[0.6ex]
k_{\ell,x} &= \frac{m_W}{2}\lf[\gamma(\cos\bbar{\vartheta} - \beta) \sin\Theta  + \sin
\bbar{\vartheta}\, \cos\bbar{\varphi}\,\cos\Theta\rg]\,,
\\[0.9ex]
k_{\ell,y} &= \frac{m_W}{2} \, \sin\bbar{\vartheta}\, \sin\bbar{\varphi}\,,
\\[0.9ex]
\begin{split}
\label{kellz}
k_{\ell,z} &= \frac{m_W}{2}[b_- \,\gamma (1 - \beta\, \cos\bbar{\vartheta}) 
\\ & \hphantom{= \frac{m_W}{2}[}+ b_+ \gamma (\cos\bbar{\vartheta} - \beta)\, \cos\Theta  -b_+\sin\bbar{\vartheta}\,
\cos\bbar{\varphi}\,\sin\Theta]\,,
\end{split}
\end{align}
where
\begin{equation}
\gamma=\frac{\sqrt{s}}{2\,m_W}\,,\qquad\beta =\sqrt{1 -
1/\gamma^2}\,,\qquad b_\pm =\frac{ 4\,E_1^2\pm
s}{4\,E_1\,\sqrt{s}}=\frac{E_1\pm E_2}{\sqrt{s}}\,.
\end{equation}
Since the jet direction ${\bf\hhat{k}_\text{jet}}$ is already defined
in the $W^-$ rest system, the relation to the angles of
Fig.~\ref{fig:conv} is,
\begin{equation}
 {\bf\hhat{k}_\text{jet}}=\lf(\begin{matrix}
 \sin\vartheta\,\cos\varphi \\
 \sin\vartheta\,\sin\varphi \\
 \cos\vartheta
\end{matrix}\rg)\,.
\label{kjet1}
\end{equation}
Eqs. (\ref{kwx}) to (\ref{kellz}) together with (\ref{kjet1}) specify
$F(\chi)$ and the calculation of the Jacobian (\ref{eq-jacob}) is now
straightforward. If $W^-$ decays into leptons and $W^+$ into quarks,
we have to make the replacements
\begin{equation}
\begin{split}
(\bbar{\vartheta}, \bbar{\varphi}) &\longrightarrow (\vartheta, \varphi)
\\
\beta &\longrightarrow -\beta
\end{split}
\end{equation}
in (\ref{kwx}) to (\ref{kellz}) and in (\ref{kjet1}).

With this we have collected all tools needed for the evaluation of
(\ref{eq-t2}) and of the integrals in (\ref{eoi}) and (\ref{cij}).

%%%%%%%%%%%%%%%%%%%%%%%%%%%%%%%%%%%%%%%%%%%%%%%%%%%%%%%%%%%%%%
\section{Results}
\label{secnum}
\setcounter{equation}{0}
\subsection{Anomalous couplings in $\gamma\gamma\to WW$}
\label{secnum1}

We now give the sensitivity to the anomalous couplings in the reaction
$\gamma\gamma \to WW$ where we allow all couplings to deviate from
zero simultaneously. These results are then compared to those obtained
in \cite{Nachtmann:2004ug} for $e^+e^- \to WW$. For the photon
collider mode we consider the c.m. energies of the initial $e^-e^-$
system as listed in the left-most column of Tab.~\ref{tabsl}. Energies
of 500 GeV and 800 GeV are planned for the ILC and higher energies are
supposed to be feasible at CLIC. The same energies are considered for
the (standard) $e^+e^-$ mode. The second column shows the integrated
luminosities that we assume for the $e^+e^-$ mode. Our calculations
for the reaction $\gamma\gamma \to WW$ are done for fixed
c.m. energies $\sqrt{s}$ of the two photon system as given in the
third column and with a Compton spectrum for each photon as described
in Sect~\ref{ssecxs2}. For the latter case the values for $\sqrt{s}$
listed in the table roughly correspond to the maxima in the photon
spectrum at the respective $e^-e^-$ energies. For the integrated
luminosity in the $\gamma\gamma$ mode we use the approximation
(cf. Sect. 4 of
\cite{Badelek:2001xb})
\begin{equation}
 L_{\gamma\gamma}=\frac{1}{3}L_{ee}\,.
\end{equation}
%%%%%%%%%%%%%%%%%%Tabular%%%%%%%%%%%%%%%%%%%%%%%%%%%%%%%%%%%%%%%%%%%%%
\begin{table}
\centering
\begin{tabular}{|c|c|c|c|c|}
\hline
\rule[-2.5mm]{0mm}{7.5mm}
$\sqrt{s_{ee}}$ & $L_{ee}$ & $\sqrt{s}$ & $N/10^7$ &
$\frac{8}{27}\, N/10^7$ \\ 
\hline
500 GeV & 500 $\text{fb}^{-1}$ & 400 GeV & 1.5 & 0.44 \\
800 GeV & 1.0 $\text{ab}^{-1}$ & 640 GeV & 3.0 & 0.89 \\
1.5 TeV & 1.5 $\text{ab}^{-1}$ & 1.2 TeV & 4.5 & 1.3 \\
3   TeV & 3.0 $\text{ab}^{-1}$ & 2.4 TeV & 9.0 & 2.7 \\
\hline
\end{tabular}
\caption{\label{tabsl} From the left to the right, c.m. energy of the $e^-e^-$ system (in the photon collider mode) or $e^+e^-$ system, luminosity in the $e^+e^-$ mode, c.m. energy $\sqrt{s}$ of the $\gamma\gamma$ system, total number of $W$ pairs in units of $10^7$ that are produced in $\gamma\gamma \to W W$ and number of $W$ pairs decaying semileptonically.}
\end{table}
%%%%%%%%%%%%%%%%%%%%%%%%%%%%%%%%%%%%%%%%%%%%%%%%%%%%%%%%%%%%%%%
Using the input values (\ref{eq-inputalpha}) - (\ref{eq-inputmw}) the
total cross section for $\gamma\gamma\to WW$ in the SM for unpolarised
photon beams is $\approx\,90\;\text{pb}$, almost constant in the
energy range we consider. The number of produced $W$ pairs is
therefore
\begin{equation}
N \approx \frac{1}{3} L_{ee}\ 90\;\text{pb}\,,
\end{equation}
which is given in the fourth column of Tab.~\ref{tabsl}. The number of
events that is used for the statistical errors on the anomalous
couplings is $N$ times the branching ratio $8/27$ for semileptonic
decays of the W bosons, shown in the right-most column. In the optimal
observables (\ref{oi}) any overall factor of the cross section,
e.g. the conversion factor $k$ in (\ref{cross-comp}), cancels. Thus
the total event rate appears in the covariance matrices of the
anomalous couplings only as the statistical factor in the denominator
of (\ref{vh}). The errors on the couplings for other total rates than
those listed in Tab.~\ref{tabsl} can therefore be easily calculated
from the numbers listed below. We use the listed values for the total
event rates $N$ both for the case with fixed $\gamma\gamma$
c.m. energy and for the case with the Compton spectrum.

We give the errors on the couplings in the presence of all other
couplings. These errors are obtained from the diagonal elements of the
covariance matrices of the anomalous couplings
\begin{equation}
\delta h_i =\sqrt{V(h)_{ii}}\,.
\label{deltahi}
\end{equation}
We also show the corresponding correlation matrices
\begin{equation}
W(h)_{ij} = \frac{V(h)_{ij}}{\sqrt{V (h)_{ii}\,V (h)_{jj}}}\,.
\label{wh}
\end{equation}
For these calculations we use the input values (\ref{eq-inputalpha}) -
(\ref{eq-inputmw}). For the Higgs mass we choose two different values
namely 120 and 150 GeV. In Tab.~\ref{tabh1} we show the sensitivities
(\ref{deltahi}) and correlation matrices (\ref{wh}) using the
covariance matrix (\ref{vh}) for various fixed $\gamma\gamma$
c.m. energies $\sqrt{s}$ and a Higgs mass of 120 GeV.  In that case we
have to deal only with the jet ambiguity, see App.~\ref{sec-kine}. We
observe that there is no correlation between $CP$-violating and
$CP$-conserving couplings. For $\sqrt{s}=400\;\text{GeV}$ all
reachable errors are between $2.3\times 10^{-4}$ and $1.2 \times
10^{-3}$, except for $h_{\tilde{W}\! B}$ where it is $3.4\times
10^{-3}$. For all couplings except for $h_{\tilde{W}\! B}$ and
$h_{\varphi\tilde{W}\!\tilde{B}}$ the sensitivity improves with rising
energy, viz. a factor 1.6 to 3.9 with each energy step. This is in
part a consequence of the increasing number $N$ of produced $W$ pairs
entering the covariance matrix (\ref{vh}), see Tab.~\ref{tabsl}. Since
the sensitivities (\ref{deltahi}) are proportional to $1/\sqrt{N}$,
this will lead to an improvement of approximately 2.5 going from the
smallest to the largest energies. We see that this is not the whole
effect. Additionally, the sensitivities increase just because the
impact of anomalous, higher-dimensional operators has to increase with
rising energy.

In contrast, the errors on $h_{\tilde{W}\! B}$ and
$h_{\varphi\tilde{W}\!\tilde{B}}$ increase slightly with rising
energy. To understand this effect one has to take a closer look on
the respective anomalous amplitudes in \cite{part1} in comparison to
the SM amplitude. Let us just mention that the matrix (\ref{cij}) and
the sensitivity would vanish if the optimal observable (\ref{oi}) were
constant even if the absolute value of (\ref{oi}) would strongly
increase. Hence the sensitivity can decrease even for increasing
anomalous contribution, if the contribution of the
corresponding dimension-six operator to the amplitude becomes
proportional to the SM amplitude at high energies.

For the $CP$ conserving couplings at 400 GeV most correlations are
about 50 \%. At higher energies only the correlation between
$h_{W\!B}$ and $h_{\varphi W\! B}$ is large whereas all others in the
CP conserving sector are about 20\% or smaller. This is different for
the CP violating couplings where all correlations are larger than
about 60\% at all energies. In particular, the correlation between
$h_{\tilde{W}\!  B}$ and $h_{\varphi\tilde{W}\! B}$ is -0.98 at 640
GeV and -1.0 at higher energies within the numerical errors. This can
be understood as follows: For longitudinally polarised $W$ bosons the
three amplitudes $\mathcal{M}_{\tilde{W}\!B}$,
$\mathcal{M}_{\varphi\tilde{W}}$ and $\mathcal{M}_{\varphi\tilde{B}}$
show in the high-energy limit $s\gg m_W^2$ the same dependence on the
photon helicities and on the angle $\Theta$, see App.~D in
\cite{part1}. Since for transversely polarised $W$ bosons the corresponding amplitudes are suppressed in this limit, 
the couplings $h_{\varphi\tilde{W}\!\tilde{B}}$ and
$h_{\tilde{W}\!B}$ are highly correlated for $s\gg m_W^2$.
Similar arguments explain the above mentioned energy behaviour of
$\delta h_{\tilde{W}\! B}$ and $\delta h_{\varphi
\tilde{W}\!\tilde{B}}$.

%%%%%%%%%%%%%%%%%%Tabular%%%%%%%%%%%%%%%%%%%%%%%%%%%%%%%%%%%%%%%%%%%%%
\begin{table}[t!]
\centering
 \begin{tabular}{|c|c|ccc||c|c|ccc|	}
\hline
    \multicolumn{5}{|c||}{$CP$-conserving couplings} & 
    \multicolumn{5}{|c|}{$CP$-violating couplings}
\\ \hline
    \multicolumn{10}{|c|}{400 GeV}
\\  \hline 
& $\delta h$ & & $W(h)$ & & & $\delta h$ & & $W(h)$& \\ \cline{3-5} \cline{8-10}
 $h$ & $\times 10^3 $ & $h_W$ & $h_{W\! B}$ & $h_{\varphi W\! B}$ & 
 $h$ & $\times 10^3 $ & $h_{\tilde{W}}$ & $h_{\tilde{W}\! B}$ & $h_{\varphi\tilde{W}\! \tilde{B}}$ \\ \hline
$h_W$ & 0.23 & 1 & 0.479 & -0.256 &
 $h_{\tilde{W}}$ & 0.31 & 1 & 0.690 & -0.556 \\
 $h_{W\! B}$ & 0.89 & 0.479 & 1 & -0.496 &
 $h_{\tilde{W}\! B}$ & 3.41 & 0.690 & 1 & -0.852 \\
 $h_{\varphi W\! B}$ & 1.16 & -0.256 & -0.496 & 1 &
 $h_{\varphi\tilde{W}\!\tilde{B}}$ & 1.13 & -0.556 & -0.852 & 1 
\\ \hline
    \multicolumn{10}{|c|}{640 GeV}
\\ \hline 
& $\delta h$ & & $W(h)$ & &  & $\delta h$ & & $W(h)$& \\ \cline{3-5} \cline{8-10}
$h$ & $\times 10^3$ & $h_W$ & $h_{W\! B}$ & $h_{\varphi W\! B}$ &
 $h$ & $\times 10^3$ & $h_{\tilde{W}}$ & $h_{\tilde{W}\! B}$ & $h_{\varphi\tilde{W}\! \tilde{B}}$ \\ \hline
$h_W$ & 0.083 & 1 & 0.332 & -0.254 &
 $h_{\tilde{W}}$ &  0.12 & 1 & 0.667 & -0.654 \\
 $h_{W\! B}$ & 0.50  & 0.332 & 1 & -0.648 &
 $h_{\tilde{W}\! B}$ & 	3.29 & 0.667  & 1 & -0.979 \\
 $h_{\varphi W\! B}$ & 0.62  & -0.254 & -0.648 & 1 &
 $h_{\varphi\tilde{W}\!\tilde{B}}$ & 1.26 & -0.654 & -0.979 & 1
 \\ \hline
    \multicolumn{10}{|c|}{1200 GeV}
\\ \hline
& $\delta h$ & & $W(h)$& &
& $\delta h$ & & $W(h)$&  \\ \cline{3-5} \cline{8-10}
$h$ & $\times 10^3$ & $h_W$ & $h_{W\! B}$ & $h_{\varphi W\! B}$ &
 $h$ & $\times 10^3$ & $h_{\tilde{W}}$ & $h_{\tilde{W}\! B}$ & $h_{\varphi\tilde{W}\! \tilde{B}}$ \\ \hline
$h_W$ &  0.033 & 1 & 0.178 & -0.167 &
 $h_{\tilde{W}}$ &   0.048 & 1 & 0.654 & -0.655 \\
$h_{W\! B}$ & 0.32 & 0.178 & 1 & -0.792 &
 $h_{\tilde{W}\! B}$ & 4.61 & 0.654 & 1 & -0.998 \\
$h_{\varphi W\! B}$ & 0.34 & -0.167 & -0.792 & 1 & 
 $h_{\varphi\tilde{W}\!\tilde{B}}$ & 1.88 & -0.655 & -0.998 & 1 
 \\ \hline
    \multicolumn{10}{|c|}{2400 GeV}
\\ \hline
& $\delta h$ & & $W(h)$ &  &
& $\delta h$ & & $W(h)$ &  \\ \cline{3-5} \cline{8-10}	
$h$ & $\times 10^3$ & $h_W$ & $h_{W\! B}$ & $h_{\varphi W\! B}$ &
 $h$ & $\times 10^3$ & $h_{\tilde{W}}$ & $h_{\tilde{W}\! B}$ & $h_{\varphi\tilde{W}\! \tilde{B}}$ \\ \hline
$h_W$ &  0.011	& 1 & 0.086 &	-0.092 &
$h_{\tilde{W}}$ & 0.015 & 1 &  0.585 & -0.586 \\
$h_{W\! B}$ & 0.18 & 0.086  &	1 &	-0.907 &
$h_{\tilde{W}\! B}$ & 5.20 & 0.585 & 1 & -1.000 \\
$h_{\varphi W\! B}$ & 0.17 & -0.092 & -0.907 & 1 &
$h_{\varphi\tilde{W}\!\tilde{B}}$  & 2.17 & -0.586 & -1.000 & 1
\\ \hline
\end{tabular}
\caption{\label{tabh1} Errors $\delta h$ in units of $10^{-3}$ for the $CP$ conserving (left) and $CP$ violating couplings (right) in the presence of all other couplings and correlation matrices $W(h)$ for a $\gamma\gamma$ c.m. energy of $\sqrt{s} = 400,\; 640,\; 1200$ and 2400 GeV with unpolarised beams. The mass of the Higgs boson is set to 120 GeV.}
\end{table}
%%%%%%%%%%%%%%%%%%%%%%%%%%%%%%%%%%%%%%%%%%%%%%%%%%%%%%%%%%%%%%%

To illustrate the dependence on the Higgs mass, we show in
Tab.~\ref{tabhhiggs} sensitivities and correlation matrices calculated
under the same conditions as those in Tab.~\ref{tabh1}. Only the Higgs
mass is increased from 120 to 150 GeV. For the $CP$-conserving
couplings only the coupling $h_{\varphi\,WB}$ is influenced. For the
smallest energy the sensitivity increases around 8\% and is
unchanged for the larger energies. The sensitivity on the
$CP$-violating coupling $h_{\tilde{W}\!B}$ increases around less then
1\%. For the $CP$-violating coupling $h_{\varphi
\tilde{W}\!\tilde{B}}$ we observe the strongest dependence on the
Higgs mass. The sensitivity increases between 13\% at the smallest
energy and around 1\% at the largest energy. In conclusion we see that the
dependence on the Higgs mass is small for the mass range 120-150
GeV. This is at present roughly the favoured mass window from direct
searches and indirect evidence \cite{Eidelman:2004wy}. In the following we will focus
on one certain value of the Higgs mass namely 120 GeV. 

%%%%%%%%%%%%%%%%%%Tabular%%%%%%%%%%%%%%%%%%%%%%%%%%%%%%%%%%%%%%%%%%%%%
\begin{table}[t!]
\centering
 \begin{tabular}{|c|c|ccc||c|c|ccc|	}
\hline
    \multicolumn{5}{|c||}{$CP$-conserving couplings} & 
    \multicolumn{5}{|c|}{$CP$-violating couplings}
\\ \hline
    \multicolumn{10}{|c|}{400 GeV}
\\  \hline 
& $\delta h$ & & $W(h)$ & & & $\delta h$ & & $W(h)$& \\ \cline{3-5} \cline{8-10}
 $h$ & $\times 10^3 $ & $h_W$ & $h_{W\! B}$ & $h_{\varphi W\! B}$ & 
 $h$ & $\times 10^3 $ & $h_{\tilde{W}}$ & $h_{\tilde{W}\! B}$ & $h_{\varphi\tilde{W}\! \tilde{B}}$ \\ \hline
$h_W$ & 0.23 & 1 & 0.477 & -0.241 &
 $h_{\tilde{W}}$ & 0.31 & 1 & 0.688 & -0.536 \\
 $h_{W\! B}$ & 0.89 & 0.477 & 1 & -0.466 &
 $h_{\tilde{W}\! B}$ & 3.39 & 0.688 & 1 & -0.829 \\
 $h_{\varphi W\! B}$ & 1.07 & -0.241 & -0.466 & 1 &
 $h_{\varphi\tilde{W}\!\tilde{B}}$ & 0.99 & -0.536 & -0.829 & 1 
\\ \hline
    \multicolumn{10}{|c|}{640 GeV}
\\ \hline 
& $\delta h$ & & $W(h)$ & &  & $\delta h$ & & $W(h)$& \\ \cline{3-5} \cline{8-10}
$h$ & $\times 10^3$ & $h_W$ & $h_{W\! B}$ & $h_{\varphi W\! B}$ &
 $h$ & $\times 10^3$ & $h_{\tilde{W}}$ & $h_{\tilde{W}\! B}$ & $h_{\varphi\tilde{W}\! \tilde{B}}$ \\ \hline
$h_W$ & 0.083 & 1 & 0.331 & -0.250 &
 $h_{\tilde{W}}$ &  0.12 & 1 & 0.671 & -0.657 \\
 $h_{W\! B}$ & 0.50  & 0.331 & 1 & -0.638 &
 $h_{\tilde{W}\! B}$ & 	3.33 & 0.671  & 1 & -0.979 \\
 $h_{\varphi W\! B}$ & 0.60  & -0.250 & -0.638 & 1 &
 $h_{\varphi\tilde{W}\!\tilde{B}}$ & 1.22 & -0.657 & -0.979 & 1
 \\ \hline
    \multicolumn{10}{|c|}{1200 GeV}
\\ \hline
& $\delta h$ & & $W(h)$& &
& $\delta h$ & & $W(h)$&  \\ \cline{3-5} \cline{8-10}
$h$ & $\times 10^3$ & $h_W$ & $h_{W\! B}$ & $h_{\varphi W\! B}$ &
 $h$ & $\times 10^3$ & $h_{\tilde{W}}$ & $h_{\tilde{W}\! B}$ & $h_{\varphi\tilde{W}\! \tilde{B}}$ \\ \hline
$h_W$ &  0.033 & 1 & 0.178 & -0.167 &
 $h_{\tilde{W}}$ &   0.048 & 1 & 0.655 & -0.655 \\
$h_{W\! B}$ & 0.32 & 0.178 & 1 & -0.790 &
 $h_{\tilde{W}\! B}$ & 4.61 & 0.655 & 1 & -0.998 \\
$h_{\varphi W\! B}$ & 0.34 & -0.167 & -0.790 & 1 & 
 $h_{\varphi\tilde{W}\!\tilde{B}}$ & 1.86 & -0.655 & -0.998 & 1 
 \\ \hline
    \multicolumn{10}{|c|}{2400 GeV}
\\ \hline
& $\delta h$ & & $W(h)$ &  &
& $\delta h$ & & $W(h)$ &  \\ \cline{3-5} \cline{8-10}	
$h$ & $\times 10^3$ & $h_W$ & $h_{W\! B}$ & $h_{\varphi W\! B}$ &
 $h$ & $\times 10^3$ & $h_{\tilde{W}}$ & $h_{\tilde{W}\! B}$ & $h_{\varphi\tilde{W}\! \tilde{B}}$ \\ \hline
$h_W$ &  0.011	& 1 & 0.086 &	-0.092 &
$h_{\tilde{W}}$ & 0.015 & 1 &  0.583 & -0.583 \\
$h_{W\! B}$ & 0.18 & 0.086  &	1 &	-0.907 &
$h_{\tilde{W}\! B}$ & 5.15 & 0.583 & 1 & -1.000 \\
$h_{\varphi W\! B}$ & 0.17 & -0.092 & -0.907 & 1 &
$h_{\varphi\tilde{W}\!\tilde{B}}$  & 2.14 & -0.583 & -1.000 & 1
\\ \hline
\end{tabular}
\caption{\label{tabhhiggs} Similar to Tab.~\ref{tabh1} but with a Higgs mass of 150 GeV.}
\end{table}
%%%%%%%%%%%%%%%%%%%%%%%%%%%%%%%%%%%%%%%%%%%%%%%%%%%%%%%%%%%%%%%

%%%%%%%%%%%%%%%%%%Tabular%%%%%%%%%%%%%%%%%%%%%%%%%%%%%%%%%%%%%%%%%%%%%
\begin{table}[t!]
\centering
 \begin{tabular}{|c|c|ccc||c|c|ccc|	}
\hline
    \multicolumn{5}{|c||}{$CP$-conserving couplings} & 
    \multicolumn{5}{|c|}{$CP$-violating couplings}
\\ \hline
    \multicolumn{10}{|c|}{500 GeV}
\\  \hline 
& $\delta h$ & & $W(h)$ & & & $\delta h$ & & $W(h)$ & \\ \cline{3-5} \cline{8-10}
$h$ & $\times 10^3$ & $h_W$ & $h_{W\! B}$ & $h_{\varphi W\! B}$ &
 $h$ & $\times 10^3$ & $h_{\tilde{W}}$ & $h_{\tilde{W}\! B}$ & $h_{\varphi\tilde{W}\! \tilde{B}}$ \\ \hline
$h_W$ & 0.36 & 1 & 0.519 & -0.120 &
$h_{\tilde{W}}$ & 0.46 & 1 & 0.630 & -0.238 \\
$h_{W\! B}$ & 1.08 & 0.519 & 1 & -0.299 &
$h_{\tilde{W}\! B}$ & 3.17 & 0.630 & 1 & -0.550 \\
$h_{\varphi W\! B}$ & 1.17 & -0.120 & -0.299 & 1 &
$h_{\varphi\tilde{W}\!\tilde{B}}$ & 1.01 & -0.238 & -0.550 & 1 \\ \hline
  \multicolumn{10}{|c|}{800 GeV}
\\  \hline 
& $\delta h$ & & $W(h)$ & & & $\delta h$ & & $W(h)$ & \\ \cline{3-5} \cline{8-10}
$h$ & $\times 10^3$ & $h_W$ & $h_{W\! B}$ & $h_{\varphi W\! B}$ &
$h$ & $\times 10^3$ & $h_{\tilde{W}}$ & $h_{\tilde{W}\! B}$ & $h_{\varphi\tilde{W}\! \tilde{B}}$ \\ \hline
$h_W$ & 0.13 & 1 & 0.407 & -0.256 &
$h_{\tilde{W}}$ &  0.17 & 1 & 0.553 & -0.491 \\
$h_{W\! B}$ & 0.60 & 0.407 & 1 & -0.547 &
$h_{\tilde{W}\! B}$ & 	2.64 & 0.553  & 1 & -0.904 \\
$h_{\varphi W\! B}$ & 0.74 & -0.256 & -0.547 & 1 & 
$h_{\varphi\tilde{W}\!\tilde{B}}$ & 0.97 & -0.491 & -0.904 & 1 \\ \hline
  \multicolumn{10}{|c|}{1500 GeV}
\\  \hline 
& $\delta h$ & & $W(h)$ &  & & $\delta h$ & & $W(h)$ & \\ \cline{3-5} \cline{8-10}
 $h$ & $\times 10^3$ & $h_W$ & $h_{W\! B}$ & $h_{\varphi W\! B}$ &
 $h$ & $\times 10^3$ & $h_{\tilde{W}}$ & $h_{\tilde{W}\! B}$ &
$h_{\varphi\tilde{W}\! \tilde{B}}$ \\ \hline
$h_W$ &  0.050	& 1 & 0.265 & -0.231 & 
$h_{\tilde{W}}$ &   0.074 & 1 & 0.528 & -0.525 \\
$h_{W\! B}$ & 0.40 & 0.265 & 1 & -0.741 &
$h_{\tilde{W}\! B}$ & 3.46 & 0.528 & 1 & -0.988 \\
$h_{\varphi W\! B}$ & 0.44 & -0.231 & -0.741 & 1 &
$h_{\varphi\tilde{W}\!\tilde{B}}$ & 1.39 & -0.525 & -0.988 & 1 \\ \hline
 \multicolumn{10}{|c|}{3000 GeV}
\\  \hline 
& $\delta h$ & & $W(h)$ & & & $\delta h$ & & $W(h)$ & \\ \cline{3-5} \cline{8-10}
 $h$ & $\times 10^3$ & $h_W$ & $h_{W\! B}$ & $h_{\varphi W\! B}$ &
 $h$ & $\times 10^3$ & $h_{\tilde{W}}$ & $h_{\tilde{W}\! B}$ & $h_{\varphi\tilde{W}\! \tilde{B}}$ \\ \hline
$h_W$ &  0.016 & 1 & 0.146 & -0.145 &
$h_{\tilde{W}}$ & 0.026 & 1 &  0.508 &  -0.509 \\
$h_{W\! B}$ & 0.23 & 0.146  &	1 & -0.881 &
$h_{\tilde{W}\! B}$ & 4.33 & 0.508  & 1 & -0.999 \\
$h_{\varphi W\! B}$ & 0.22 & -0.145 &  -0.881 & 1 &
$h_{\varphi\tilde{W}\!\tilde{B}}$  & 1.80 & -0.509 & -0.999 & 1
\\ \hline
\end{tabular}
\caption{\label{tabh2} Errors $\delta h_i$ and correlation matrices
$W(h)$ in units of $10^{-3}$ for the $CP$ conserving (left) and $CP$
violating couplings (right) in the presence of all other
couplings. We consider photons obtained through Compton
backscattering off electrons with an $e e$ c.m. energy of
$\sqrt{s_{ee}} = 500$, 800, 1500 and 3000 GeV. The mass of the Higgs boson is set to 120 GeV.
The photons are supposed to be unpolarised and have an energy distributed according to a Compton
spectrum, see Sect.~\ref{ssecxs2}. Since approximately 80\% of the
c.m. energy will be transfered into the $\gamma\gamma$ system these
results are comparable to the results from Tab.~\ref{tabh1}.}
\end{table}
%%%%%%%%%%%%%%%%%%%%%%%%%%%%%%%%%%%%%%%%%%%%%%%%%%%%%%%%%%%%%%%

The results listed in Tab.~\ref{tabh2} are similar to those in
Tab.~\ref{tabh1}, but here the photons are distributed according to
the Compton spectrum (CS). The relevant differential cross section is now
given by (\ref{stilde}).  Furthermore, in addition to the jet
ambiguity the neutrino ambiguity enters the calculation, see
App.~\ref{sec-kine}. As we can see, if we compare the
Tabs.~\ref{tabh1} and \ref{tabh2}, the results do not change so
much. Due to the additional ambiguity, the errors for $h_W$, $h_{W\! 
B}$, $h_{\varphi W\!B}$ and $h_{\tilde{W}}$ are slightly higher. In
contrast, the errors for $h_{\tilde{W}\! B}$, $h_{\varphi
\tilde{W}\!\tilde{B}}$ are smaller except for the 3000 GeV case. This is easily understood: As
already discussed above, the errors for these two couplings decrease
with decreasing fixed $\gamma\gamma$ c.m. energies. Taking the Compton
spectrum into account, also lower energies with a better sensitivity
will now enter the final result.

\subsection{Comparison with $e^+e^- \to WW$}
\label{secnum2}

Finally, we would like to compare our results from Sect.~\ref{secnum1}
with those obtained for the reaction $e^+e^- \to WW$ in
\cite{Nachtmann:2004ug}. Tab.~\ref{tabhfin} combines the sensitivities
reachable at a $e^+e^-$ collider \cite{Nachtmann:2004ug} and a
$\gamma\gamma$ collider with the bounds that can be obtained using
present data \cite{Nachtmann:2004ug}. We see that the
couplings $h_{\varphi W\! B}$ and $h_{\varphi \tilde{W}\!\tilde{B}}$ are only testable at a
$\gamma\gamma$ collider since only here the anomalous $\gamma\gamma H$
vertex enters, see Tab.~\ref{tab:vertices} and \cite{part1}. This
already underlines the importance of the $\gamma\gamma$ mode at a
future LC. Concerning the reachable sensitivities for the couplings
which are testable in both modes, the differences are quite small. Only for
the couplings $h_{W\!B}$ and $h_{\tilde{W}\!B}$ one sees a clear
tendency of the $e^+e^-$ mode to give the better sensitivities.

On the left hand side of Tab.~\ref{tabhfin} we list the constraints
\cite{Nachtmann:2004ug} on the anomalous couplings that we can get
from present data. These data cover precision observables
and results of the direct measurement of triple gauge-couplings in
$e^+e^-\to WW$ at LEP2. Surprisingly, we get good constraints on
$h_\varphi^{(3)}$ and $h_{W\!B}$ from present high-precision
observables, see Sect.~\ref{sec-lep} and \cite{Nachtmann:2004ug}. The
sensitivity for $h_{W\!B}$ reachable in $W$-pair production at a
future LC of the next generation is only of the same order as the
present bound. Only at a LC with an even larger luminosity and energy
like CLIC~\cite{Ellis:1998wx} we expect an improvement for this
particular coupling. For the coupling $h_\varphi^{(3)}$ we can not expect
any improvements through $W$-pair production. The best way to improve
the knowledge about these two couplings at a LC would be to decrease
the errors of the precision observables further. The Giga-$Z$ mode
offers such a possibility. A
measurement at the $Z$~pole with an event rate that is about 100~times
that of LEP1 should in essence reduce the errors~$\delta h_{W\!B}$ and
$\delta h_\varphi^{(3)}$ given in Tab.~\ref{tabhfin} by a factor of 10.
The errors on the other couplings which are constrained only
by the direct measurements of the triple gauge-couplings will improve
significantly at a future LC.

%%%%%%%%%%%%%%%%%%Tabular%%%%%%%%%%%%%%%%%%%%%%%%%%%%%%%%%%%%%%%%%%%%%
\begin{table}[th!]
\centering
\begin{tabular}{|c|c|c||c|c|c|c|c|c|}
\hline
 & \multicolumn{2}{|c||}{} & \multicolumn{6}{|c|}{Sensitivity at a LC} \\ \cline{4-9} 
 & \multicolumn{2}{|c||}{Constraints from} &  \multicolumn{2}{|c}{$e^+e^-$}
 & \multicolumn{2}{|c}{$\gamma\gamma$ mode}  & \multicolumn{2}{|c|}{$\gamma\gamma$ mode} \\
 & \multicolumn{2}{|c||}{LEP  and SLD}  & \multicolumn{2}{|c}{mode}
 & \multicolumn{2}{|c}{fixed $\sqrt{s_{\gamma\gamma}}$}  & \multicolumn{2}{|c|}{ with CS} \\ \cline{2-9}
& $m_H$ & $h_i$  & $\sqrt{s_{ee}}$ & $\delta h_i$ & $\sqrt{s_{\gamma\gamma}}$ & $\delta h_i$ & $\sqrt{s_{ee}}$ & $\delta h_i$ \\
 & [GeV]  & $\times 10^3$ &[GeV]  & $\times 10^3$ &[GeV] & $\times 10^3$ & [GeV] & $\times 10^3$  \\ \hline
 \multicolumn{9}{|c|}{Measureable $CP$-conserving couplings} \\ \hline
%%%%%%%%%%%%%%%%%%%%%%%%%%%%%%%%%%%%%%%%%%%%%%%%%%%%%%%%%%%%%%%%%%%%%%%%%%%%%%%%%%%%%%%%%%%%%%%%%%%%%
% h_W
%%%%%%%%%%%%%%%%%%%%%%%%%%%%%%%%%%%%%%%%%%%%%%%%%%%%%%%%%%%%%%%%%%%%%%%%%%%%%%%%%%%%%%%%%%%%%%%%%%%%%
& \multicolumn{2}{c||}{$-69\pm 	39$}  & 500  & 0.28  & 400  & 0.23  & 500  & 0.36 \\
\raisebox{-1.ex}[0ex][0ex]{$h_W$}
& \multicolumn{2}{c||}{Constraint from}  & 800  & 0.12  & 640  & 0.083 & 800  & 0.13 \\
& \multicolumn{2}{c||}{TGCs measurement}  &      &       & 1200 & 0.033 & 1500 & 0.050 \\
& \multicolumn{2}{c||}{at LEP 2}          & 3000 & 0.018 & 2400 & 0.011 & 3000 & 0.016 \\ \hline

%& 120 & $-62.4\pm 36.3$ & 500  & 0.28  & 400  & 0.23  & 500  & 0.36  \\
%$h_W$
%& 200 & $-62.5\pm 36.3$ & 800  & 0.12  & 640  & 0.083 & 800  & 0.13  \\
%& 500 & $-62.8\pm 36.3$ &      &       & 1200 & 0.033 & 1500 & 0.050 \\
%&     &               & 3000   & 0.018 & 2400 & 0.011 & 3000 & 0.016 \\ \hline
%%%%%%%%%%%%%%%%%%%%%%%%%%%%%%%%%%%%%%%%%%%%%%%%%%%%%%%%%%%%%%%%%%%%%%%%%%%%%%%%%%%%%%%%%%%%%%%%%%%%%
% h_WB
%%%%%%%%%%%%%%%%%%%%%%%%%%%%%%%%%%%%%%%%%%%%%%%%%%%%%%%%%%%%%%%%%%%%%%%%%%%%%%%%%%%%%%%%%%%%%%%%%%%%%
& \raisebox{-1.4ex}[0ex][0ex]{120} & \raisebox{-1.4ex}[0ex][0ex]{$-0.06\pm 0.79$} & 500  & 0.32  & 400  & 0.89 & 500  & 1.08 \\
\raisebox{-1.4ex}[0ex][0ex]{$h_{W\!B} $}
& \raisebox{-1.4ex}[0ex][0ex]{200} & \raisebox{-1.4ex}[0ex][0ex]{$-0.22\pm 0.79$} & 800  & 0.16  & 640  & 0.50 & 800  & 0.60 \\
& \raisebox{-1.4ex}[0ex][0ex]{500} & \raisebox{-1.4ex}[0ex][0ex]{$-0.45\pm 0.79$} &      &       & 1200 & 0.32 & 1500 & 0.40 \\
&     &                 & 3000 & 0.015 & 2400 & 0.18 & 3000 & 0.23 \\ \hline
%%%%%%%%%%%%%%%%%%%%%%%%%%%%%%%%%%%%%%%%%%%%%%%%%%%%%%%%%%%%%%%%%%%%%%%%%%%%%%%%%%%%%%%%%%%%%%%%%%%%%
% h_phiWB
%%%%%%%%%%%%%%%%%%%%%%%%%%%%%%%%%%%%%%%%%%%%%%%%%%%%%%%%%%%%%%%%%%%%%%%%%%%%%%%%%%%%%%%%%%%%%%%%%%%%%
& \multicolumn{2}{c||}{}  & \multicolumn{2}{c|}{}  & 400  & 1.16 & 500  & 1.17 \\
\raisebox{-1.4ex}[0ex][0ex]{$h_{\varphi W\!B} $}
& \multicolumn{2}{c||}{Does not}  & \multicolumn{2}{c|}{Does not}  & 640  & 0.62 & 800  & 0.74 \\
& \multicolumn{2}{c||}{contribute}  & \multicolumn{2}{c|}{contribute}  & 1200 & 0.34 & 1500 & 0.44  \\
& \multicolumn{2}{c||}{}  & \multicolumn{2}{c|}{}  & 2400 & 0.17 & 3000 & 0.22 \\ \hline
%%%%%%%%%%%%%%%%%%%%%%%%%%%%%%%%%%%%%%%%%%%%%%%%%%%%%%%%%%%%%%%%%%%%%%%%%%%%%%%%%%%%%%%%%%%%%%%%%%%%%
% h_\varphi^(3)
%%%%%%%%%%%%%%%%%%%%%%%%%%%%%%%%%%%%%%%%%%%%%%%%%%%%%%%%%%%%%%%%%%%%%%%%%%%%%%%%%%%%%%%%%%%%%%%%%%%%%
& \raisebox{-1.4ex}[0ex][0ex]{120} & \raisebox{-1.4ex}[0ex][0ex]{$-1.15\pm 2.39$} & 500  & 36.4 & \multicolumn{4}{c|}{}     \\
\raisebox{-1.4ex}[0ex][0ex]{$h_{\varphi}^{(3)} $}
& \raisebox{-1.4ex}[0ex][0ex]{200} & \raisebox{-1.4ex}[0ex][0ex]{$-1.86\pm 2.39$} & 800  & 53.7 & \multicolumn{4}{c|}{Does not}     \\
& \raisebox{-1.4ex}[0ex][0ex]{500} & \raisebox{-1.4ex}[0ex][0ex]{$-3.79\pm 2.39$} &    &      & \multicolumn{4}{c|}{contribute}      \\
&     &                 & 3000 &  ``$\infty$''    & \multicolumn{4}{c|}{} \\ \hline
 \multicolumn{9}{|c|}{Measureable $CP$-violating couplings} \\ \hline
%%%%%%%%%%%%%%%%%%%%%%%%%%%%%%%%%%%%%%%%%%%%%%%%%%%%%%%%%%%%%%%%%%%%%%%%%%%%%%%%%%%%%%%%%%%%%%%%%%%%%
% h_tW
%%%%%%%%%%%%%%%%%%%%%%%%%%%%%%%%%%%%%%%%%%%%%%%%%%%%%%%%%%%%%%%%%%%%%%%%%%%%%%%%%%%%%%%%%%%%%%%%%%%%%
& \multicolumn{2}{c||}{$68\pm 	81$}  & 500  & 0.28  & 400  & 0.31 & 500  & 0.46 \\
\raisebox{-1.4ex}[0ex][0ex]{$h_{\tilde{W}}$}
& \multicolumn{2}{c||}{Constraint from}  & 800  & 0.12  & 640  & 0.12 & 800  & 0.17 \\
& \multicolumn{2}{c||}{TGCs measurement}  &      &       & 1200 & 0.048 & 1500 & 0.074 \\
& \multicolumn{2}{c||}{at LEP 2}  & 3000 & 0.018 & 2400 & 0.015 & 3000 & 0.026 \\ \hline
%%%%%%%%%%%%%%%%%%%%%%%%%%%%%%%%%%%%%%%%%%%%%%%%%%%%%%%%%%%%%%%%%%%%%%%%%%%%%%%%%%%%%%%%%%%%%%%%%%%%%
% h_tWB
%%%%%%%%%%%%%%%%%%%%%%%%%%%%%%%%%%%%%%%%%%%%%%%%%%%%%%%%%%%%%%%%%%%%%%%%%%%%%%%%%%%%%%%%%%%%%%%%%%%%%
& \multicolumn{2}{c||}{$33\pm 84$}  & 500  & 2.2  & 400  & 3.41 & 500  & 3.17 \\
\raisebox{-1.4ex}[0ex][0ex]{$h_{\tilde{W}\!B}$}
& \multicolumn{2}{c||}{Constraint from}  & 800  & 1.4  & 640  & 3.29 & 800  & 2.64 \\
& \multicolumn{2}{c||}{TGCs measurement}  &      &      & 1200 & 4.61 & 1500 & 3.46 \\
& \multicolumn{2}{c||}{at LEP 2}  & 3000 & 0.77 & 2400 & 5.20 & 3000 & 4.33 \\ \hline
%%%%%%%%%%%%%%%%%%%%%%%%%%%%%%%%%%%%%%%%%%%%%%%%%%%%%%%%%%%%%%%%%%%%%%%%%%%%%%%%%%%%%%%%%%%%%%%%%%%%%
% h_phi tWtB
%%%%%%%%%%%%%%%%%%%%%%%%%%%%%%%%%%%%%%%%%%%%%%%%%%%%%%%%%%%%%%%%%%%%%%%%%%%%%%%%%%%%%%%%%%%%%%%%%%%%%
& \multicolumn{2}{c||}{}  &  \multicolumn{2}{c|}{} & 400  & 1.13 & 500  & 1.01 \\
\raisebox{-1.4ex}[0ex][0ex]{$h_{\varphi \tilde{W}\!\tilde{B}}$}
& \multicolumn{2}{c||}{Does not}  & \multicolumn{2}{c|}{Does not} & 640  & 1.26 & 800  & 0.97 \\
& \multicolumn{2}{c||}{contribute}  & \multicolumn{2}{c|}{contribute}      & 1200 & 1.88 & 1500 & 1.39 \\
& \multicolumn{2}{c||}{}  & \multicolumn{2}{c|}{}      & 2400 & 2.17 & 3000 & 1.80 \\ \hline
\end{tabular}

\caption{\label{tabhfin} 
The present constraints from LEP and SLD as calculated in \cite{Nachtmann:2004ug}
and reviewed in Sect.~\ref{sec-lep} and the expected sensitivities
reachable in the different modes at a future LC are shown. We assume
the integrated luminosities and the number of $W$ pairs produced in
$\gamma\gamma\to WW$ as given in Tab.~\ref{tabsl}. For the calculation
of the reachable sensitivity at a LC we choose a Higgs mass of 120 GeV.}
\end{table}
%%%%%%%%%%%%%%%%%%%%%%%%%%%%%%%%%%%%%%%%%%%%%%%%%%%%%%%%%%%%%%%

%%%%%%%%%%%%%%%%%%%%%%%%%%%%%%%%%%%%%%%%%%%%%%%%%%%%%%%%%%%%%%%%%
%%%%%%%%%%%%%%%%%%%%%%%%%%%%%%%%%%%%%%%%%%%%%%%%%%%%%%%%%%%%%%%%%

\section{Conclusions}
\label{sec-conc}
\setcounter{equation}{0}

%%%%%%%%%%%%%%%%%%%%%%%%%%%%Conclusions from gamma1%%%%%%%%%%%%%%%%%%%%%%%%%%%%

We have presented an analysis of the phenomenology of the gauge-boson
sector of an electroweak effective Lagrangian that is locally
$SU(2)\times U(1)$ invariant. In addition to the SM~Lagrangian we
included all ten dimension-six operators that are built either only
from the gauge-boson fields of the~SM or from the gauge-boson fields
combined with the SM-Higgs field.

In a preceding work~\cite{Nachtmann:2004ug} the impact of these
anomalous couplings onto observables from $Z$ decays and $W$
production at hadron and $e^+e^-$ colliders were studied in this
framework.  For a large class of observables the anomalous effects
only show up through a modified effective leptonic weak mixing
angle. Other observables depend on the anomalous couplings in a
different way and therefore lead to further constraints.  From all
data constraints on three $CP$~conserving and two $CP$~violating
couplings were derived as reviewed in Sect.~\ref{sec-lep}.

In the present paper we calculated the statistically best possible
bounds on the anomalous couplings that can be obtained from
$\gamma\gamma\to WW$ at a photon collider by means of optimal
observables. The couplings $h_W, h_{W\! B}, h_{\tilde{W}}$ and
$h_{\tilde{W}\! B}$ can be measured both in \mbox{$\gamma
\gamma\rightarrow WW$} and in \mbox{$e^+e^- \rightarrow WW$}. The
sensitivity to these anomalous couplings achievable in the two reactions
is similar. The couplings $h_{\varphi W\!B}$ and
$h_{\varphi\tilde{W}\!\tilde{B}}$ can only be measured in
$\gamma\gamma\to WW$.

We point out that one gets already today constraints for $h_{W\!B}$ and
$h_\varphi^{(3)}$ from precision observables which are quite
comparable with the sensitivity that one expects to reach in $e^+e^-\to
WW$ and $\gamma\gamma\to WW$ at an ILC, see
Sect.~\ref{secnum2}. Hence, we expect to get the best constraints for
these two couplings by improving the accuracy of the
precision observables, e.g. in the Giga-$Z$ mode of an ILC.

We summarise our findings. An ILC with $e^+e^-$ collisions at
$\sqrt{s}=500\;\text{GeV}$ will improve the sensitivity to the
couplings $h_W,\;h_{\tilde{W}}$ and $h_{\tilde{W}\!B}$ compared to the
present bounds by factors of about 140, 290 and 38, respectively. The
Giga $Z$ option will improve the bounds on $h_{W\!B}$ and
$h_{\varphi}^{(3)}$ by about a factor of 10. Only the $\gamma\gamma$
collider will make the study of the couplings $h_{\varphi W\!B}$ and
$h_{\varphi \tilde{W}\!\tilde{B}}$ possible. The obtainable
sensitivities are comparable to those for the other couplings from the
$e^+e^-$ mode. Three combinations out of the original ten anomalous
couplings in (\ref{eq-Leff2}), that is $h_{\varphi}^{(1)},\;
h^\prime_{\varphi W\!B}$ (\ref{eq-hphiWBprime}) and $h^\prime_{\varphi
\tilde{W}\!\tilde{B}}$ (\ref{eq-hphiWtildeBprime}) remain unmeasurable from the
normalised event distributions of the reactions considered here.

A quantitative analysis of the sensitivities to anomalous couplings as
presented here should help to decide how much total luminosity is
required in each mode of a future ILC. As already explained in
Sect.~\ref{sec-intro}, our approach, using the effective Lagrangian
(\ref{eq-Leff}) instead of form factors, is perfectly suited for a
comprehensive study of all constraints on the $h_i$ coming from
different modes at an ILC and from high precision observables. We have
seen that in any case the
\mbox{$e^+ e^-$} and the \mbox{$\gamma\gamma$} modes deliver complementary constraints on the anomalous
couplings of the effective Lagrangian considered. Both modes as well
as the Giga-$Z$ mode are indispensable for a comprehensive study of
the gauge-boson sector at a future~ILC.

%%%%%%%%%%%%%%%%%%%%%%%%%%%%%%%%%%%%%%%%%%%%%%%%%%%%%%%%%%%%%%%%%
%%%%%%%%%%%%%%%%%%%%%%%%%%%%%%%%%%%%%%%%%%%%%%%%%%%%%%%%%%%%%%%%%

\section*{Acknowledgements}

The authors are grateful to M.~Diehl for reading a draft of this
manuscript and to A.~Denner and A.~de~Roeck for useful discussions.
This work was supported by the German Bundesministerium f\"ur Bildung
und Forschung, BMBF project no. 05HT4VHA/0, and the Deutsche
Forschungsgemeinschaft through the Graduierten\-kolleg ``Physikalische
Systeme mit vielen Freiheitsgraden''.

%%%%%%%%%%%%%%%%%%%%%%%%%%%%%%%%%%%%%%%%%%%%%%%%%%%%%%%%%%%%%%%%%%%%%%

\begin{appendix}

\section{Reconstruction ambiguities at a photon collider}
\label{sec-kine}
\setcounter{equation}{0}

In Sect.~\ref{sec-opti} we discussed the consequences of reconstruction
ambiguities for the calculation of the optimal
observables (\ref{oi}) and the covariance matrix (\ref{vh}). Here we discuss
the two ambiguities appearing at a photon collider, that is the one
from the incomplete knowledge of the neutrino momentum and the one
from no tag for the quark and antiquark jets.

The first ambiguity enters only if we consider a Compton spectrum for
the photon energies. The lack of a direct measurement of the neutrino
energy and momentum leads to a two-fold ambiguity in the
identification of $E_1$ and $\sqrt{s}$. Remember that the variables in
(\ref{xi}) are the observable ones. Let ${\bf k}_\nu$ be the neutrino momentum
in the LS. Its component perpendicular to the beam axis is given by
\begin{equation}
{\bf k}_{\nu,\perp}=-{\bf k}_{\ell,\perp}-{\bf k}_{W,\perp}\,,
\end{equation}
with ${\bf k}_{W,\perp}=(k_{W,x},0)$. If
${\bf k}_{\ell,\perp}\neq 0$ we have
\begin{equation}
k_{\nu,z}=\frac{1}{{\bf k}_{\ell,\perp}^2}\lf(r\,k_{\ell,z}\pm
g\,E_\ell\rg)\,, \qquad E_{\nu}=\frac{1}{{\bf
k}_{\ell,\perp}^2}\lf(r\,E_\ell\pm g\,k_{\ell,z}\rg)\,,
\label{knuenu}
\end{equation}
where
\begin{equation}
r=
\frac{m_W^2}{2}+{\bf k}_{\ell\perp}\cdot{\bf k}_{\nu\perp}\,,
\qquad
g=\sqrt{r^2-{\bf k}_{\ell\perp}^2\,{\bf k}_{\nu\perp}^2}\,.
\end{equation}	
In (\ref{knuenu}) one has to {\em simultaneously} choose the upper
{\em or} lower signs, i.e. there are two corresponding solutions,
provided that both $g$ and $k_{\ell z}$ are different from zero. The
energies of the photons in the LS are obtained from energy and
momentum conservation:
\begin{align}
E_1 + E_2 &= E_W + E_\ell + E_\nu\,,
\label{E1pE2}
\\
\label{E1mE2}
E_1 - E_2 &= k_{W,z} + k_{\ell,z} + k_{\nu,z}\,.
\end{align}
For $E_1$ we have
\begin{equation}
E_1 = \frac{1}{2}(E_W + E_\ell + E_\nu + k_{W,z} + k_{\ell,z} + k_{\nu,z})\,. \label{E1EW}
\end{equation}
One can easily check that in the case where there are two solutions
(\ref{knuenu}) for $k_{\nu,z}$ and $E_\nu$, these always lead to two
different values for $E_1$. From (\ref{E1pE2}) and (\ref{E1mE2}) we
obtain the  squared $\gamma\gamma$ c.m. energy
\begin{equation}
\begin{split}
s = 4E_1E_2 &= k^2_{W,x} + {\bf k}^2_{\ell,\perp}+{\bf k}^2_{\nu,\perp}
+ 2 (E_W E_\ell - k_{W,z}k_{\ell,z}) \\ &\quad
 + 2 (E_W + E_\ell) E_\nu + 2 (k_{W,z} +
k_{\ell,z})k_{\nu,z}\,. \label{srel}
\end{split}
\end{equation} 
Inserting (\ref{knuenu}) into (\ref{srel}) we obtain in general two
solutions for $\sqrt{s}$. In some regions of the parameter space one
of these solutions must be discarded since the value of $\sqrt{s}$ is
unphysical. For some other cases where ${\bf k}_{\ell,\perp}$ or $g$
are zero the two solutions are identical. Because of these two reasons
the two-fold neutrino ambiguity disappears in part of the parameter space.

The second ambiguity appears also in the case where the photon
energies are fixed and arises since we supposed no identification of the
charge of a jet. The relation between the measurable jet direction
${\bf\hhat{k}_\text{jet}}$ and the angles $\vartheta$ and $\varphi$ is
given in (\ref{kjet1}). Since ${\bf\hhat{k}_\text{jet}}$ is already
defined in the $W^-$ ($W^+$) rest frame, the second jet appearing in
the hadronic decay of the $W$ boson points in the opposite
direction. Hence it is clear that the lack of jet-charge tagging leads
to a two-fold ambiguity in the angles $\vartheta, \varphi$.  Since the
Jacobian $J_k$ in (\ref{eq-t2}) is trivial in this case one can handle
this ambiguity basically through the restriction on one hemisphere in
(\ref{kjet1}) to avoid the double counting of indistinguishable jets.

Due to the two described cases, the variables $\chi$ (\ref{chi}) can
be reconstructed only with a four-fold or a two-fold ambiguity
depending on the input values for the measurable variables $\xi$.

\end{appendix}

%%%%%%%%%%%%%%%%%%%%%%%%%%%%%%%%%%%%%%%%%%%%%%%%%%%%%%%%%%%%%%%%%%%%%%

\end{document}